\documentclass[fleqn,usenatbib]{mnras}

\usepackage{newtxtext,newtxmath}

\usepackage[T1]{fontenc}
\usepackage{calligra}
\usepackage[normalem]{ulem}
\DeclareRobustCommand{\VAN}[3]{#2}
\let\VANthebibliography\thebibliography
\def\thebibliography{\DeclareRobustCommand{\VAN}[3]{##3}\VANthebibliography}

\usepackage{graphicx}	
\usepackage{amsmath}	

\newcommand{\orcid}[1]{\href{https://orcid.org/#1}{\includegraphics[width=10pt]{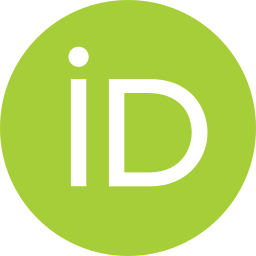}}}

\title[CRs and cooling of the low-metallicity ISM]{The impact of cosmic ray heating on the cooling of the low-metallicity interstellar medium}

\author[V. Brugaletta]{Vittoria Brugaletta$^{\orcid{0000-0003-1221-8771}}$$^{1, 2}$\thanks{E-mail: brugaletta@ph1.uni-koeln.de} , Stefanie Walch$^{1}$, Thorsten Naab$^{2}$, Philipp Girichidis$^{3}$, Tim-Eric Rathjen$^{1}$,
\newauthor Daniel Seifried$^{1}$, Pierre Colin Nürnberger$^{1}$, Richard Wünsch$^{4}$ and Simon C. O. Glover$^{3}$ 
\\
$^{1}$ I. Physikalisches Institut, Universität zu Köln, Zülpicher Str. 77, 50937 Köln, Germany \\
$^{2}$ Max Planck Institute for Astrophysics, Karl-Schwarzschild-Str. 1, 85748 Garching, Germany\\
$^{3}$ Universit\"at Heidelberg, Zentrum f\"ur Astronomie, Institut f\"ur Theoretische Astrophysik, Albert-Ueberle-Str. 2, D-69120 Heidelberg, Germany\\
$^{4}$ Astronomical Institute of the Czech Academy of Sciences, Boční II 1401, 141 00 Prague, Czech Republic\\
}

\date{Accepted 2024 December 20. Received 2024 December 20; in original form 2024 October 24}

\pubyear{2025}

\begin{document}
\label{firstpage}
\pagerange{\pageref{firstpage}--\pageref{lastpage}}
\maketitle

\begin{abstract}
Low-metallicity environments are subject to inefficient cooling. They also have low dust-to-gas ratios and therefore less efficient photoelectric (PE) heating than in solar-neighbourhood conditions, where PE heating is one of the most important heating processes in the warm neutral interstellar medium (ISM). We perform magneto-hydrodynamic simulations of stratified ISM patches with a gas metallicity of 0.02~Z$_\odot$ as part of the SILCC project. The simulations include non-equilibrium chemistry, heating, and cooling of the low-temperature ISM as well as anisotropic cosmic ray (CR) transport, and stellar tracks. We include stellar feedback in the form of far-UV and ionising (FUV and EUV) radiation, massive star winds, supernovae, and CR injection. From the local CR energy density, we compute a CR heating rate that is variable in space and time. In this way, we can compare the relative impact of PE and CR heating on the metal-poor ISM and find that CR heating can dominate over PE heating. Models with a uniform CR ionisation rate of $\zeta$~=~3~$\times$~10$^{-17}$~s$^{-1}$ suppress or severely delay star formation, since they provide a larger amount of energy to the ISM due to CR heating. Models with a variable CR ionisation rate form stars predominantly in pristine regions with low PE heating and CR ionisation rates where the metal-poor gas is able to cool efficiently. Because of the low metallicity, the amount of formed stars in all runs is not enough to trigger outflows of gas from the mid-plane. 
\end{abstract}




\section{Introduction}
Cosmic rays (CR) are highly energetic charged particles, mainly protons and electrons, but also heavier nuclei, that can travel within the interstellar medium (ISM) and influence both its dynamics and chemistry. At lower energies, below 200~MeV, electrons dominate the composition, whereas at higher energies protons dominate \citep{Cummings2016}. Early studies \citep{Baade1934, Ginzburg1964} suggest that supernova (SN) remnants in the Milky Way are the main contributors in accelerating GeV - TeV Galactic CRs via diffusive shock acceleration \citep{Krymskii1977,Axford1978,Bell1978a,Bell1978b,BlandfordOstriker1978}. At lower energies other acceleration mechanisms are important, in particular shocks in the ISM (see e.g. a review from \citealt{Hanasz2021, PadovaniEtAl2020, Ruszkowski2023} and \citealt{Meyer2024}). The CR energy spectrum spans more than ten orders of magnitude, and around 32 orders of magnitude in flux \citep{Swordy2001}. The main contribution to the energy density of CRs is provided by protons with energies of a few GeV, whereas for higher energies the energy spectrum declines as a broken power law \citep[e.g.][]{Ruszkowski2023}. The energy density of CRs measured at Earth was estimated by \citealt{Webber1998} to be around 1.8~eV~cm$^{-3}$, which is comparable to the thermal, kinetic, and magnetic energy densities measured in the Milky Way \citep{Boulares1990, Cox2005}. This approximate equipartition suggests that CRs are important for the dynamics and evolution of the ISM, and, possibly, also for the regulation of star formation.  

CRs travel through the ISM via advection with the gas, anisotropic diffusion and streaming along the magnetic field. Due to their reduced cooling efficiency, they are capable of retaining a substantial energy density over extended periods. However, due to their rapid diffusion, local variations in CR energy density are rapidly smoothed out. The CR pressure gradient is typically much smaller than the thermal pressure gradient. Consequently, SNe are able to locally shape the structure of the ISM, whereas the effect of CRs is visible only on larger temporal and spatial scales \citep{girichidis2018, silcc7}. However, the CR pressure gradient is able to slowly lift the gas from the disc \citep{girichidis2016,girichidis2018, silcc7}, supporting galactic outflows even during periods of low star formation activity \citep{Hanasz2013, silcc7}.

However, low-energy CRs up to 1~GeV play a key role in the chemistry of the ISM because they are responsible for the ionisation and heating of the gas \citep{Field1969}, substantially impacting its evolution. In particular, low-energy CRs are able to penetrate deeply into dense clouds, to a depth where the UV radiation emitted by stars is hindered by dust absorption. Consequently, they are an important source of heating and ionisation for the cold and dense gas (see e.g.\ \citealt{Caselli1998, Bergin2007, PadovaniEtAl2020}). The CR ionisation rate of atomic hydrogen was estimated by \citealt{Spitzer1968} to be 6.8~$\times$~$10^{-18}$~s$^{-1}$ in the proximity of Earth, where solar modulation affects the measurement of the low-energy ($<$~1~GeV) CR spectrum. In fact, the interaction of low-energy CRs with the magnetised solar wind causes their deflection, preventing them from reaching the vicinity of Earth. On the other hand, more energetic CRs (few GeV at least) can reach Earth without being affected. The low-energy CR energy spectrum has been measured without the effects of solar modulation by the probes Voyager 1 \citep{Cummings2016} and Voyager 2 \citep{Stone2019} once they have passed the heliopause, after which the low-energy CRs are out of range to interact with the turbulent, magnetised solar wind. In more recent times, values of the CR ionisation rate in the range 10$^{-18}$~--~10$^{-16}$~s$^{-1}$ \citep{Sabatini2020, Sabatini2023, Socci2024} and up to 10$^{-14.5}$~s$^{-1}$ \citep{Pineda2024} have been measured for the molecular hydrogen in star-forming regions, and in the range 10$^{-16}$ -- 10$^{-14}$~s$^{-1}$ for nearby protostars \citep{Ceccarelli2014, Fontani2017, Favre2018, Cabedo2023}. This variation, both in space and time is expected to lead to a locally changing CR heating rate. 

The role of CR heating, due to the interaction of low-energy CRs with the ISM, is of primary importance in  low-metallicity environments, and will be analysed in detail in this work. At a temperature below few 10$^{6}$~K, the cooling of the ISM is dominated by metals. In metal-poor environments, the ISM therefore cools much less efficiently. These environments therefore tend to be warmer than environments in solar-metallicity conditions (Brugaletta et al., in prep.). It is therefore even more important to carefully consider the heating mechanisms that play a role in such environments. Previous works of the SILCC collaboration \citep{silcc1, silcc2,silcc3, silcc4, silcc5, silcc6, silcc7} have analysed the solar-metallicity ISM employing a constant value for the interstellar radiation field (ISRF) strength, parameterised using $G_0$ in Habing units \citep{habing1968}. Only lately \citep{silcc8} they have adopted an on-the-fly calculation of the far-ultraviolet (FUV) interstellar radiation field, which is dependent on the present stellar population. Since the PE heating rate scales linearly with the dust-to-gas ratio, it becomes comparable to the CR heating rate in low metallicity gas (see Sec.~\ref{sec:crir_method}). The latter depends linearly on the CR ionisation rate $\zeta$, which has been assumed to be constant in the previous SILCC studies except for a control run in \citet{girichidis2018}. In this paper we introduce a novel method that scales the CR ionisation rate from the spatially and temporally variable CR energy density that is already computed within our code. To test this implementation, we adopt a metallicity of 0.02~Z$_\odot$, for which we expect to observe a strong impact. Such low metallicities are measured in some local dwarf galaxies, e.g.\ the blue compact dwarf galaxy I~Zw~18 \citep[][]{Zwicky1966, French1980}, but are also seen in the early phases of the Universe (\citealt{Heintz2023, Vanzella2023, Curti2024}).

This paper is organised as follows. In Sec.~\ref{sec:numerical_methods} the numerical methods and the simulation setup are explained, including a description of the new treatment to compute the CR ionisation rate from the energy density of CRs that is already computed in our code. In Sec.~\ref{sec:global_evolution} a qualitative description of the behaviour of our simulations is provided. Sec.~\ref{sec:results} illustrates the results of our work, followed by a discussion (Sec.~\ref{sec:discussion}) and our summary and conclusions (Sec.~\ref{sec:conclusions}). 

\section{Numerical methods and simulation setup}
\label{sec:numerical_methods}

In our simulations, we employ the same setup as the SILCC framework (\citealt{silcc1, silcc2, silcc3, silcc4, silcc5, silcc6, silcc7, silcc8}). We model the evolution of the gas in as stratified disc by solving the ideal magneto-hydrodynamic equations with the three-dimensional, adaptive mesh refinement code \textsc{Flash} version 4.6 (\citealt{Fryxell_2000, Dubey_2008,dubey2009}). 

Our computational domain is an elongated box of size 500~pc $\times$ 500~pc $\times$ $\pm$ 4~kpc. Near the mid-plane, we adopt a resolution $\Delta$$x$ $\sim$ 3.9~pc, whereas for $|z| >$ 1~kpc the resolution can reach up to 7.8~pc. In the beginning of each run, the spatial distribution of the gas along the $z$-direction follows a Gaussian centred in the mid-plane with a scale height of 30~pc, whereas in the x-y plane the gas distribution is uniform. The gas surface density, $\Sigma_{\mathrm{gas}}$, is set to be 10~M$_\odot$~pc$^{-2}$ for consistency with previous SILCC works. However, this value could be an upper limit for low-mass dwarf galaxies \citep{Jaiswal2020}, or too low to describe the central regions of compact blue dwarf galaxies like I~Zw~18 \citep{Lelli2012}. We have periodic boundary conditions in the $x$- and $y$-directions, and outflow conditions in the $z$-direction. This means that the gas is able to exit the simulation box in the $z$-direction but it is not able to fall back in. To avoid a sudden gravitational collapse and, consequently, a starburst in the very beginning of the simulation, we mix the gas via turbulence driving. Kinetic energy is injected for the first 20~Myr at large scales corresponding to the size of the box (500~pc), such that the mass-weighted root mean square velocity of the gas is initially equal to 10~km~s$^{-1}$.    

 Using a tree-based method \citep{wunsch2018}, we consider four different contributions to gravity: $(i)$ the gas self-gravity; $(ii)$ the gravity due to the presence of sink particles which represent star clusters (\citealt{Dinnbier2020}, see below); $(iii)$ an external gravitational potential to mimic the presence of an old stellar population; $(iv)$ a constant dark matter potential. A more in-depth treatment of gravitational effects can be found in \citealt{silcc1, silcc3}. 
 
 All runs have a magnetic field initially oriented along the $x$-axis which changes in the vertical direction as
\begin{equation}
B_x(z) = B_{x,0} \sqrt{\rho(z)/\rho(z = 0)},     
\end{equation}
where $B_{x,0}$ = 6~$\mu$G and $\rho(z)$ is the density at a height $z$ from the mid-plane. 

To model the star formation due to gravitational collapse happening in high-density regions, we include collisionless and accreting star-cluster-sink particles (\citealt{bate1995, Federrath_2010, silcc3, Dinnbier2020}). A sink particle is created in a cell when specific criteria are met. These include a threshold density $\rho_\mathrm{thr}$~=~2~$\times~$10$^{-21}$~g~cm$^{-3}$, and the gas must be in a converging flow, Jeans-unstable, and in a gravitational potential minimum. The sink particles are responsible for the injection of momentum and thermal energy in the ISM in the form of radiation, stellar winds, supernova explosions, and the injection of CRs that are accelerated in SN shocks. One massive star is formed inside a cluster for every 120~M$_\odot$ of gas accreted onto that cluster. The initial mass of the new massive star is sampled from an IMF with a Salpeter-like slope \citep[][]{salpeter1955} in the range 9-120~M$_\odot$. We assume that the remaining amount of gas forms lower mass stars inside the cluster, which are not considered individually. We follow the time evolution of massive stars by means of the \textsc{BoOST} models \citep[][]{brott2011, szecsi2020}, considering their wind velocity, bolometric luminosity, mass loss rate and effective temperatures. At the end of their life, all massive stars are assumed to explode as Type II supernovae, injecting thermal energy or momentum depending on whether the Sedov-Taylor phase is resolved \citep{gatto2015}. In fact, if the radius at the end of the Sedov-Taylor phase is resolved with at least three cells, thermal energy of 10$^{51}$~erg is injected at the explosion site. This region is defined as the volume of the sphere centred on the supernova and with a three-cells radius ($\sim$ 11.7~pc in physical space). However, if the ambient density is too high to resolve the radius at the end of the Sedov-Taylor phase ($n_\text{crit}$ = 3.3~cm$^{-3}$, see \citealt{silcc6}), the expected momentum that the gas would have if the supernova explosion were resolved is injected instead \citep[see][]{silcc1}. In both cases the mass of the progenitor is added to the mass of the gas that was present in the injection region before the explosion. 

Feedback from radiation originating in stellar sources plays an important role in our simulations, influencing the chemistry of the ISM and, in the case of ionising photons, creating H~II regions. To treat the interaction of extreme ultra-violet (EUV) photons with the ISM, we include the radiative transfer module \textsc{TreeRay} \citep{wunsch2021}, which uses the On-The-Spot approximation \citep{Osterbrock_1988}. All photons with an energy higher than 13.6~eV are able to ionise atomic hydrogen and are treated with \textsc{TreeRay/OnTheSpot}. We compute the fraction of ionising photons assuming each star to be a black body whose spectrum can be computed by means of the star's effective temperature, taken from the stellar models. At every timestep, ionising photons emitted by all stars in a sink particle are injected in the cell where the sink is located. The \textsc{TreeRay} module uses the octal-tree already employed in the gravity solver \citep{wunsch2018} to propagate the radiation in the computational domain using a backwards radiative transfer method. For every cell in the computational domain, \textsc{TreeRay} creates a \textsc{HEALPix} sphere \citep{gorski2005} with (in this case) 48 rays cast from the centre of each target cell and oriented normal to the surface of the \textsc{HEALPix} sphere. The radiation transport equation is solved along each ray for each cell.

Important in our simulations is the implementation of heating and cooling processes, computed on--the--fly using a chemical network (\citealt{Nelson_1997}, \citealt{glover2007a, Glover_2007b}) based on the calculation of non-equilibrium abundances of seven chemical species, namely H, H$^{+}$, H$_2$, CO, C$^{+}$, O and free electrons. We assume that all the H$_2$ is formed on the surface of dust grains, following the prescription from \citet{Hollenbach1989}, since this channel dominates over the H$_2$ formation via the H$^{-}$ ion, even at the low metallicities considered in this study \citep{Glover2003}. It also dominates over the three-body channel, which becomes important for high gas densities ($n$~$>$~10$^{8}$~cm$^{-3}$, \citealt{Palla1983}) and much lower metallicities ($Z$~$<$~10$^{-6}$~Z$_\odot$, \citealt{Omukai2005}) than those treated here. H$_2$ is mainly destroyed because of photodissociation by the interstellar radiation field, however, also cosmic-ray ionization and collisional dissociation in the hot gas are taken into account. In our treatment we compute the photodissociation rate of H$_2$ following 
\begin{equation}
    R_\mathrm{pd} = R_\mathrm{pd,H_2, thin} f_\mathrm{dust, H_2} f_\mathrm{shield,H_2},
\end{equation}
where
\begin{equation}
    R_\mathrm{pd,H_2,thin} = 3.3 \times 10^{-11} \sum_i \Big( G_\mathrm{cluster, \textit{i}} R^{-2}_{i} \Big)  \  \mathrm{s}^{-1},
\end{equation}
is the photodissociation rate of H$_2$ in the optically thin gas \citep{DraineBertoldi1996}, and $G_\mathrm{cluster, \textit{i}} R^{-2}_{i}$ is the geometrically attenuated ISRF of clusters (see below). The factor  $f_\mathrm{dust,H_2}$~=~exp(-4.18~$\times$~A$_\mathrm{V, 3D}$), whose exponent is taken from \citealt{Heays2017}, with the local visual extinction A$_\mathrm{V, 3D}$ determined by \textsc{TreeRay/OpticalDepth} \citep{wunsch2018}, accounts for the effects of dust extinction encountered by the radiation that photodissociates H$_2$ molecules. The visual extinction is computed as \citep{Bohlin1978}
\begin{equation}
    A_\mathrm{V, 3D} = N_\mathrm{H,tot} /(1.87\times10^{21}~\mathrm{cm}^{-2}) \times \ Z,
    \label{eq:Av}
\end{equation}
with the local 3D-averaged column density $N_\mathrm{H,tot}$ obtained from \textsc{TreeRay/OpticalDepth} \citep{wunsch2018}, and $Z$ the metallicity. The factor $f_\mathrm{shield,H_2}$ accounts for the effects of H$_2$ self-shielding.  Regarding heating processes, we include, among others, PE heating by dust grains and heating due to CRs. We describe them in more detail in Sec.~\ref{sec:heating_rates}. Concerning the cooling of the gas with temperatures higher than 10$^4$~K, we assume collisional ionisation equilibrium for helium and metals, and employ the tabulated cooling rates from \citealt{Gnat2012}. In this regime, the Lyman-$\alpha$ cooling is computed using the non-equilibrium abundance of atomic hydrogen that is already computed in our chemistry network. We do not include chemical enrichment due to stellar feedback.

\subsection{The \textsc{AdaptiveG0} module}
\label{sec:AdaptiveG0}
We utilise an updated and self-consistent approach to modelling the FUV radiation field ($E_\gamma$ = 6 - 13.6~eV), as introduced in \citealt{silcc8}. Departing from a static ISRF solely subject to local dust attenuation (using \textsc{TreeRay/OpticalDepth}, see below), we now incorporate the intensity of the FUV radiation field from all formed star clusters individually. We first compute the FUV luminosity for each star cluster i, $G_\mathrm{cluster, i}$, given its mass and age, using \textsc{Starburst99} single-stellar population synthesis models \citep{Leitherer1999}. Then we apply a geometrical attenuation and sum over all clusters 
to obtain $G_0$ for each cell:
\begin{equation}
G_\mathrm{0} = \sum_i G_\mathrm{cluster, i} \times R^{-2}_{i} \, .
\end{equation}
Here, $R_{i}$ is the distance from the considered cell to the $i$-th cluster. Next,  we apply a minimum floor $G_\mathrm{bg}$~=~0.0948 to the calculated $G_0$ as done in \citealt{silcc8}. This background value is computed for a cosmic UV background from \citet{HaardtMadau2012} plus the contribution of a static low-mass stellar population with a stellar surface density of $\Sigma_\star$~=~30~M$_\odot$~pc$^{-2}$. 

Finally, (as for the runs with constant $G_0$), we apply the local extinction by dust attenuation to obtain an effective ISRF at each cell, $G_\mathrm{eff}$:
\begin{equation}
    G_\mathrm{eff} = G_0 \times \mathrm{exp}(-2.5 \ A_\mathrm{V, 3D}) \, .
\end{equation}

\subsection{Cosmic rays}
We include CRs in the form of a separate relativistic fluid within an advection-diffusion approximation. Therefore, we solve modified MHD equations, e.g. including an additional CR energy source term $Q_\text{cr}$ \citep{girichidis2016, girichidis2018}. The equations read
\begin{equation}
    \frac{\partial \rho}{\partial t} + \nabla \cdot (\rho \boldsymbol{v}) = 0,
\end{equation}
\begin{equation}
    \frac{\partial \rho \boldsymbol{v}}{\partial t} + \nabla \cdot \Big( \rho \boldsymbol{v} \boldsymbol{v}^T - \frac{\boldsymbol{B} \boldsymbol{B}^T}{4 \pi} \Big) + \nabla P_\mathrm{tot} = \rho \boldsymbol{g} + \dot{\boldsymbol{q}}_\mathrm{sn},
 \end{equation}

\begin{align}
    \nonumber  \frac{\partial e}{\partial t} + & \nabla \cdot \Big[\ (e + P_\mathrm{tot})\boldsymbol{v} - \frac{\boldsymbol{B} (\boldsymbol{B} \cdot \boldsymbol{v})}{4\pi} \Big]\  \\ & = \rho \boldsymbol{v} \cdot \boldsymbol{g} + \nabla \cdot (\boldsymbol{\mathrm{K}} \nabla e_\mathrm{cr}) + \dot{u}_\mathrm{chem} + \dot{u}_\mathrm{sn} + Q_\mathrm{cr},
\end{align}

\begin{equation}
    \frac{\partial \boldsymbol{B}}{\partial t} - \nabla \times (\boldsymbol{v} \times \boldsymbol{B}) = 0,
\end{equation}
\begin{equation}
    \frac{\partial e_\mathrm{cr}}{\partial t} + \nabla \times (e_\mathrm{cr} \boldsymbol{v}) = - P_\mathrm{cr} \nabla \cdot \boldsymbol{v} + \nabla \cdot (\boldsymbol{\mathrm{K}} \nabla e_\mathrm{cr}) + Q_\mathrm{cr},
\end{equation}
where $\rho$ is the density, $\boldsymbol{v}$ is the velocity of the gas, $\boldsymbol{B}$ the magnetic field, $P_\text{tot} = P_\text{thermal} + P_\text{kinetic} + P_\text{cr}$ the total pressure, $\boldsymbol{g}$ the gravitational acceleration, $\boldsymbol{\dot{q}}_\text{sn}$ the momentum input of unresolved SNe, $e = \frac{\rho v^2}{2} + e_\text{thermal} + e_\text{cr} + \frac{B^2}{8 \pi}$ the total energy density, $\boldsymbol{K}$ the CR diffusion tensor, $\dot{u}_\text{chem}$ the change in thermal energy due to heating and cooling processes, $\dot{u}_\text{sn}$ the thermal energy input from resolved supernovae, $Q_\text{cr} = Q_\text{cr, injection} + \Lambda_\text{hadronic}$. The term $Q_\text{cr}$ accounts for the injection of 10$^{50}$~erg of energy per supernova explosion in form of CRs (see e.g. \citealt{Hillas2005, Ackermann2013}), and the cooling of CRs via hadronic and adiabatic losses (see e.g. \citealt{Pfrommer2017, Girichidis2020}). For the latter we follow \citealt{Pfrommer2017} assuming
\begin{equation}
    \Lambda_\mathrm{hadronic} = -7.44 \times 10^{-16} \times \Big( \frac{n_e}{\mathrm{cm}^{-3}} \Big) \times \Big( \frac{e_\mathrm{cr}}{\mathrm{erg \ cm}^{-3}} \Big) \  \mathrm{erg} \ \mathrm{s}^{-1}\mathrm{cm}^{-3}. 
\end{equation}

For the diffusion tensor we adopt two constant components fo the diffusion coefficient, $K_\parallel$~=~10$^{28}$~cm$^2$~s$^{-1}$ and $K_\perp$~=~10$^{26}$~cm$^2$~s$^{-1}$ parallel and perpendicular to the magnetic field lines, respectively \citep{Strong2007, Nava2013}. CR diffusion is solved in an operator split manner. We evolve the modified MHD equations on a hydrodynamical time step, computed via a modified effective sound speed using ($P_\mathrm{th+cr} = P_\mathrm{th} + P_\mathrm{cr}$). The diffusion is solved with an explicit solver in sub-cycling with a local diffusion time step for each sub-cycle 
\begin{equation}
    \Delta t_\mathrm{diff} = \min\left(\Delta t_\mathrm{hydro}, 0.5\,C_\mathrm{CFL}\frac{(\Delta x)^2}{K_\parallel+K_\perp}\right). 
\end{equation}

\subsection{Photoelectric and CR heating rates}
\label{sec:heating_rates}
Low-metallicity environments cool less efficiently due to the lack of metals, the main coolants of the ISM for temperatures below few 10$^6$~K (see e.g \citealt{Wolfire1995, Bialy2019}). Therefore, the ISM is more sensitive to the different heating processes. The two most important heating mechanisms in this case are the PE heating, which is the main heating mechanism in the warm ISM, also at higher metallicities, and the CR heating. 

In the local ISM, the heating rate for photoelectric heating $\Gamma_\mathrm{pe}$ is given by \citep{Bakes1994, Bergin2004}
\begin{equation}
    \Gamma_\mathrm{pe} = 1.3 \times 10^{-24} \epsilon G_\mathrm{eff} n d\ \ \ \ \ [ \text{erg s}^{-1} \text{cm}^{-3}],
    \label{eq:PE_rate}
\end{equation}
where $n$ is the number density of hydrogen nuclei, $d$ the dust-to-gas mass ratio in per cent, with $d$~=~1 (one per cent) in solar-neighbourhood conditions. The PE heating efficiency $\epsilon$ reads \citep{Bakes1994, Wolfire2003}
\begin{equation}
\label{eq:pe_eff}
    \epsilon = \frac{0.049}{1 + (\psi/963)^{0.73}} +  \frac{0.037 (T/10^4)^{0.7}}{1 + (\psi/2500)},
\end{equation}
with 
\begin{equation}
    \psi = \frac{G_\mathrm{eff} T^{0.5}}{n_e},
\end{equation}
where $T$ is the temperature of the gas, and $n_e$ the electron number density. This expression for the PE heating efficiency was derived for conditions comparable to those in the local ISM. For simplicity we assume that it also holds in the low metallicity environment studied in this paper. In practice, this is unlikely to be the case: low metallicity galaxies are strongly deficient in polycyclic aromatic hydrocarbons (PAHs; see e.g.\ \citealt{draine2007}; \citealt{Sandstrom2012}), which make a substantial contribution to the total PE heating rate. It is therefore likely that the true PE heating efficiency in the very metal-poor ISM will be significantly smaller than the value given by Equation~\ref{eq:pe_eff}, which would render PE heating even less effective than we find in this study.

The CR heating rate we adopt follows \citet{goldsmith1978} assuming that each ionisation deposits 20~eV as heat, and reads
\begin{align}
   \nonumber \Gamma_\mathrm{cr} & =  \ 20 \ \zeta \ (n_\mathrm{H_2} + n_\mathrm{H}) \ \ \ \ \ [\text{eV s}^{-1}\text{cm}^{-3}]   \\ & = 3.2 \times 10^{-11} \zeta (n_\mathrm{H_2} + n_\mathrm{H}) \ \ \ \ \ [ \text{erg s}^{-1} \text{cm}^{-3}],
\label{eq:CRheating}
\end{align}
where $\zeta$ is the CR ionisation rate expressed in units of s$^{-1}$, $n_\mathrm{H_2}$ the number density of H$_2$ and $n_\mathrm{H}$ the number density of H. 

In previous works (e.g. \citealt{silcc6, silcc7}), both the parameters $G_0$ for the PE heating and $\zeta$ for the CR heating rates have been assumed to be constant and in the range $G_0$~=~1.7 -- 42.7 (with $G_0$~=~1.7 valid for solar-neighbourhood conditions; \citealt{draine1978}) and in the range $\zeta =3 \times$~10$^{-17}$ -- 3~$\times$~10$^{-16}$~s$^{-1}$ depending on the adopted gas surface density, respectively. In this case, the PE and CR heating rates were already variable, but assuming constant values of $G_0$ and $\zeta$. Therefore, a PE heating similar to that of solar-neighbourhood environments has been assumed even in the absence of massive stars, as well as CR heating in the absence of supernovae accelerating galactic CRs. We show that these conditions deeply affect the possibility of stars forming in the metal-poor ISM (see Sec.~\ref{sec:results}).

\begin{figure}
	\includegraphics[width=\columnwidth]{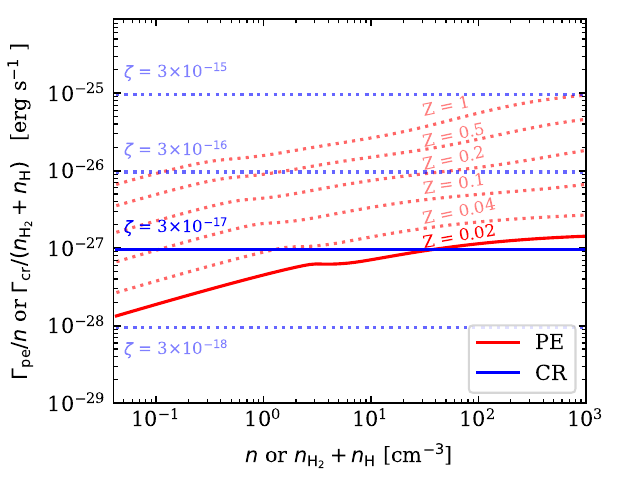}
    \caption{Ratio of the photoelectric heating rate (PE, red) with the number density of hydrogen nuclei, and ratio of the CR heating rate (CR, blue) with the sum $(n_\mathrm{H_2} + n_\mathrm{H})$, following Eq.~\ref{eq:CRheating}. Both heating rates are computed for a number of different metallicities, ranging from 0.02~Z$_\odot$ to 1~Z$_\odot$, using our chemistry module in standalone mode. The PE heating rate has been computed for optically thin conditions assuming $G_0$~=~1.7, and it varies with metallicity as it scales linearly with the dust-to-gas ratio, which is a function of metallicity. The CR heating rate has been computed also for a number of different CR ionisation rates ($\zeta$), as it does not depend on metallicity. The blue solid line highlights the heating rate with $\zeta$~=~3~$\times$~10$^{-17}$~s$^{-1}$, which is the value assumed in previous works of the SILCC collaboration for a gas surface density of 10~M$_\odot$~pc$^{-2}$. The red solid line marks the PE heating rate computed at the metallicity relevant for this work, assuming a linear scaling of the dust-to-gas ratio with metallicity. We note that in this case the cosmic-ray heating rate is similar, if not more important, than the PE heating rate at high densities.}
    \label{fig:constant_PE_CR}
\end{figure}

In Fig.~\ref{fig:constant_PE_CR} we show both $\Gamma_\mathrm{pe}$ and $\Gamma_\mathrm{cr}$ divided by the number density of hydrogen nuclei and the sum $n_\mathrm{H_2} + n_\mathrm{H}$, respectively. They are computed for different metallicities ranging from 1~Z$_\odot$ down to 0.02~Z$_\odot$. We compute these values with our chemistry module in standalone mode for 1~Gyr, after which the gas has reached chemical and thermal equilibrium. The rates were computed assuming a constant $G_0 = 1.7$, optically thin gas with $A_\mathrm{V, 3D}$~=~0,  and four different constant values of $\zeta$. Regarding the PE heating, we note that the rate computed for solar metallicity is around two orders of magnitude higher than the same rate at a metallicity of 0.02~Z$_\odot$. This is due to the fact that, for this specific calculation, we are scaling the dust-to-gas ratio linearly with metallicity, meaning that $d$~=~0.02 per cent at 0.02~Z$_\odot$. Therefore, at low metallicity, PE heating plays a less important role in the heating of the gas compared to solar-neighbourhood conditions. Concerning the CR heating, we see that the rate depends on the value of $\zeta$, and hence is metallicity-invariant. Therefore, even without any shielding by dust, PE and CR heating rates are comparable in the range of densities relevant in our low-metallicity simulations, meaning that the way we model CR heating can strongly affect our results.

\subsection{Implementation of a variable cosmic ray ionisation rate}
\label{sec:crir_method}
 In this work, we aim at modelling the CR ionisation rate, $\zeta$, consistently with the CR energy density, $e_\mathrm{cr}$, that is already computed by our code. As seen in the previous section, for every supernova explosion we inject 10$^{50}$~erg of energy in the form of CRs, which are treated in the fluid approximation. 
 
  To compute $\zeta$ from $e_\mathrm{cr}$, we consider the local total hydrogen column density, $N_\mathrm{H,tot}$, which is an average over all 48 directions computed by \textsc{TreeRay/OpticalDepth}. We use $N_\mathrm{H, tot}$ instead of the H$_2$ column density, $N_\mathrm{H_2}$, as reported by \citet{Padovani2009, Padovani2022}. This is due to the small amount of H$_2$ at the low metallicity we consider, therefore we would greatly underestimate the CR attenuation if we were using only the H$_2$ column density. In the following, we will nevertheless apply the prescription from \citet{Padovani2009, Padovani2022} obtained for proton impact on the H$_2$ gas, since no analogous relation for the atomic gas is provided. With this choice we speculate our attenuation of $\zeta$ to be affected by an uncertainty of around a factor of 2 \citep{Glassgold1974}, compared to the attenuation shown in Fig.~C1 in \citealt{Padovani2022}. 
  
  We define a threshold column density $N_\mathrm{H, thres}$~=~10$^{20}$~cm$^{-2}$, above which we attenuate the value of $\zeta$. The choice of this value is based on Fig.~C.1 from \citealt{Padovani2022}, where the attenuation of $\zeta$ starts for $N_\mathrm{H_2}$~=~10$^{20}$~cm$^{-2}$. Since no data is available below this threshold, we assume no attenuation in this range. We therefore compute $\zeta$ from $e_\mathrm{cr}$ as in the following.
\begin{enumerate}
    \item In the more diffuse gas where $N_\mathrm{H, tot}$ < $N_\mathrm{H, thres}$, we scale $\zeta$ linearly with $e_\mathrm{cr}$ according to 
\begin{equation}
\label{eq:cr_scale}
\zeta = 3 \times 10^{-17} \left(\frac{e_{\rm cr}}{1 \: {\rm eV \, cm^{-3}}} \right) \: {\rm s^{-1}}.
\end{equation} 
    This is to connect the value 3~$\times$~10$^{-17}$~s$^{-1}$ of the CR ionisation rate used in previous SILCC works with the $e_\mathrm{cr}$ that is on average observed in runs with $\Sigma_\mathrm{gas}$~=~10~M$_\odot$~pc$^{-2}$, of around 1~eV~cm$^{-3}$ \citep{silcc6}.
    \item  In the denser gas where $N_\mathrm{H, tot}$ > $N_\mathrm{H, thres}$, we multiply the linear scaling given in Equation~\ref{eq:cr_scale} with an additional attenuation factor, $c_\mathrm{att}$, defined as
    \begin{equation}
        c_\mathrm{att} = (N_\mathrm{H, tot}/N_\mathrm{H, thres})^{-0.423},
        \label{eq:cr_att}
    \end{equation}
    where the exponent -0.423 is given by the prescription for the spectrum of protons and heavy nuclei in \citet{Padovani2009}, which is reasonable for diffuse clouds. This is physically motivated by the idea that a low energy CR travelling in a very dense medium will interact several times with the atoms and molecules of that medium, losing part of its energy and therefore its capability of ionising new atoms. We verified that this exponent is well in accordance with the average slope, computed in logarithmic scale, that we obtain from Fig. C.1 in \citealt{Padovani2022} considering the two most extreme models shown there, their $\mathcal{L}$~model and their model with power law index $\alpha$~=~-1.2.
\end{enumerate}

As initial conditions for our simulations we set $e_\mathrm{cr}$, and therefore $\zeta$, to zero. The CRs relevant for the energy computation are those injected from supernova explosions and the only source of CRs would be those coming from nearby galaxies. However, the CRs able to travel for such long distances (more than 500~pc, the length of our box) are few in number and have a cross section for collisions with the gas which is very small \citep{Padovani2022}. Hence, we assume their contribution to the CR energy density to be negligible.

\subsection{Simulation parameters}
\label{sec:simulation_parameters}
In this work, we run four simulations, as listed in Table~\ref{tab:simulations}, where we vary our prescription for the PE and CR heating. In one of our simulations (Z0.02), we assume the presence of a constant interstellar radiation field (ISRF), whose strength is expressed in units of the Habing field \citep{habing1968}, $G_0$ = 1.7. This value is valid for solar-neighbourhood conditions \citep{draine1978}, therefore we could expect this value to be not optimal for very metal-poor environments. In run Z0.02, as in all our simulations, we attenuate the ISRF according to the local dust attentuation, which is determined from the local total column density $N_\mathrm{H,tot}$ \citep[see][]{silcc1}. Overall, the setup of Z0.02 is the same as in \citealt{silcc7}, except for the different metallicity and different stellar tracks.  

All the other simulations employ the novel \textsc{AdaptiveG0} module from \citealt{silcc8}, described in Sec.~\ref{sec:AdaptiveG0}, which computes the value of $G_\mathrm{eff}$ consistently from the distribution of young star clusters. Regarding the CR heating, a constant value of the CR ionisation rate $\zeta$~=~3~$\times$~10$^{-17}$~s$^{-1}$ is assumed for runs Z0.02 and Z0.02-vG$_0$, whereas for the two simulations Z0.02-vG$_0$-v$\zeta$ and Z0.02-vG$_0$-v$\zeta$-BS, $\zeta$ is computed using the scaling explained in Sec.~\ref{sec:crir_method}. In this way, we aim to explore the impact of the variability of the ISRF and/or of $\zeta$ on the PE and CR heating rates and the resulting properties of the low-metallicity ISM. 
The only difference between the Z0.02-vG$_0$-v$\zeta$ and Z0.02-vG$_0$-v$\zeta$-BS runs is the scaling of the dust-to-gas ratio with metallicity. In all simulations but Z0.02-vG$_0$-v$\zeta$-BS, the dust-to-gas ratio is scaled linearly with the metallicity of the run, assuming $d$~=~1 (one percent) at solar metallicity. Recent observations of low-metallicity galaxies suggest that the scaling of the dust-to-gas ratio with metallicity becomes steeper than linear in low-metallicity regimes \citep{Remy-Ruyer2014}. Therefore, we analyse the impact of a different scaling in the Z0.02-vG$_0$-v$\zeta$-BS run, where we adopt the power-law scaling from \citealt{Bialy2019}, which holds for metallicities lower than 0.2~Z$_\odot$, and reads
\begin{equation}
    d = Z’_0 (Z'/Z'_0)^\alpha,
\end{equation}
where $Z'_0$~=~0.2 and $Z'$~=~0.02 is the metallicity of the gas and $\alpha$~=~3. This results in a value $d$~$\approx$~10$^{-4}$ in percentage.

\begin{table*}
	\centering
	\caption{List of simulations performed in this work (first column). $G_0$ (second column) describes the strength of the ISRF in Habing units \citep{habing1968} and $\zeta$ (third column) is the CR ionisation rate in s$^{-1}$. In the fourth column the scaling of the dust-to-gas ratio ($d$) with metallicity is specified. In the last column we give $t_\mathrm{SF}$, which is the simulation time at which the first star cluster has been formed, if star formation takes place.}
	\label{tab:simulations}
	\begin{tabular}{lcccr} 
		\hline
		Simulation & $G_0$ & $\zeta$ [s$^{-1}$]& $d$ scaling & $t_\mathrm{SF}$ [Myr]\\
		\hline
        Z0.02 & 1.7 & $ 3~\times~10^{-17}$ & Linear & No star formation\\
        Z0.02-vG$_0$ & Variable & $ 3~\times~10^{-17}$ & Linear & 128.2\\
        Z0.02-vG$_0$-v$\zeta$ & Variable & Variable & Linear & 44.2\\
        Z0.02-vG$_0$-v$\zeta$-BS & Variable &  Variable & \citealt{Bialy2019} & 44.5\\
        \hline
	\end{tabular}
\end{table*}

\section{Results}
\label{sec:results}
\subsection{Global evolution}
\label{sec:global_evolution}

\begin{figure*}
	\includegraphics[width=0.9\textwidth]{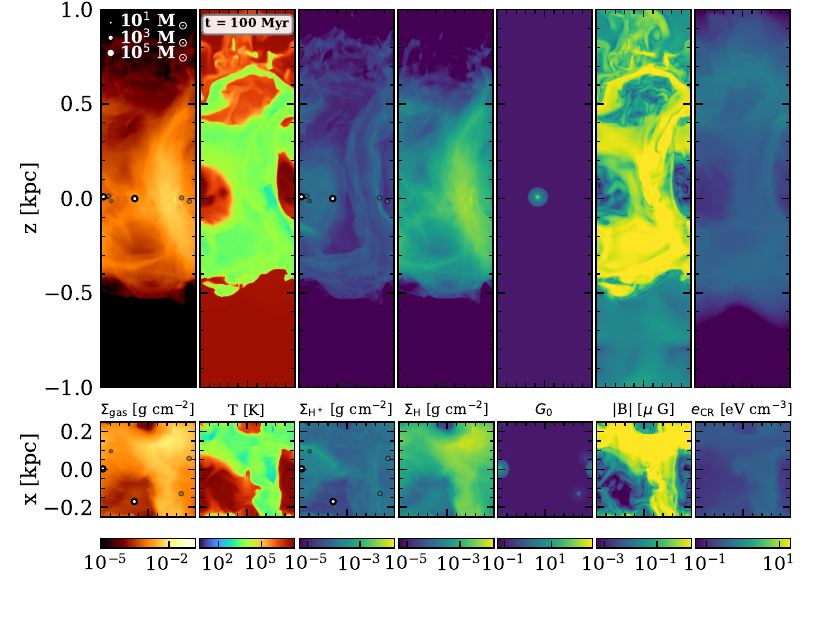}
    \caption{Snapshot of the Z0.02-vG$_0$-v$\zeta$-BS run at $t$~=~100~Myr (simulation time). The top row shows the edge-on view and the bottom row the face-on view of the following quantities: the column density of the gas $\Sigma_\mathrm{gas}$, a temperature slice, the column density of H$^+$, H, a slice of the variable G$_0$, a slice of the amplitude of the magnetic field $B$, a slice of the energy density of CRs $e_\mathrm{cr}$, respectively. In the vertical direction, the box is not shown in its entirety as it is cut for clarity reasons at $\pm$~1~kpc. Slices are taken at $y=0$ (top row) or $z=0$ (bottom row). The white circles represent the active star clusters (not in scale with the figure), whereas the transparent circles represent the inactive star clusters. }
\label{fig:snapshot}
\end{figure*}

 We show a representative snapshot of run Z0.02-vG$_0$-v$\zeta$-BS in Fig.~\ref{fig:snapshot}. The initial conditions for our simulations are described in detail in \citealt{silcc1}. In the beginning of every run the gas is mixed by the initial turbulence driving, while it is slowly collapsing towards the mid-plane because of cooling and gravity. During this phase the gas can form dense and cold regions where stars are born. The time when the sink formation criteria are fulfilled greatly depends on the individual run. 
 
 In our simulations all stars are spawned in clusters and start their evolution directly on the main sequence, injecting energy and momentum into the ISM because of their radiation, stellar winds and supernovae. This results in an increase in the ionised hydrogen abundance in the regions close to the newly–born clusters, becoming more and more extended as the clusters accrete, forming more new stars. During their life, star clusters can accrete gas as well as lose gas because of stellar winds.  

At the end of their life, all the massive stars explode as Type II supernovae, whose overall effect is to heat the gas and to push it in the vertical direction, making the scale-height of the disc thicker. Since we follow star formation self-consistently, and do not use a fixed star formation rate, more intense periods of star formation are followed by those of lower star-forming activity. As a consequence, when the majority of stars have already exploded and no new stars are being formed, the outflow-driving force exerted on the gas weakens. Therefore, the gas starts to fall back down towards the mid-plane. When a certain amount of gas has already reached the original disc, the density is high enough to create new molecular clouds, which in turn form new star clusters (see e.g. \citealt{silcc6}). After stars are born, the gas is pushed outwards by feedback until the majority of the stars are not active anymore. Then, it will fall down toward the mid-plane creating new star clusters, restarting the cycle.

\subsection{The spatial and temporal variability of $\zeta$}
\begin{figure}
	\includegraphics[width=\columnwidth]{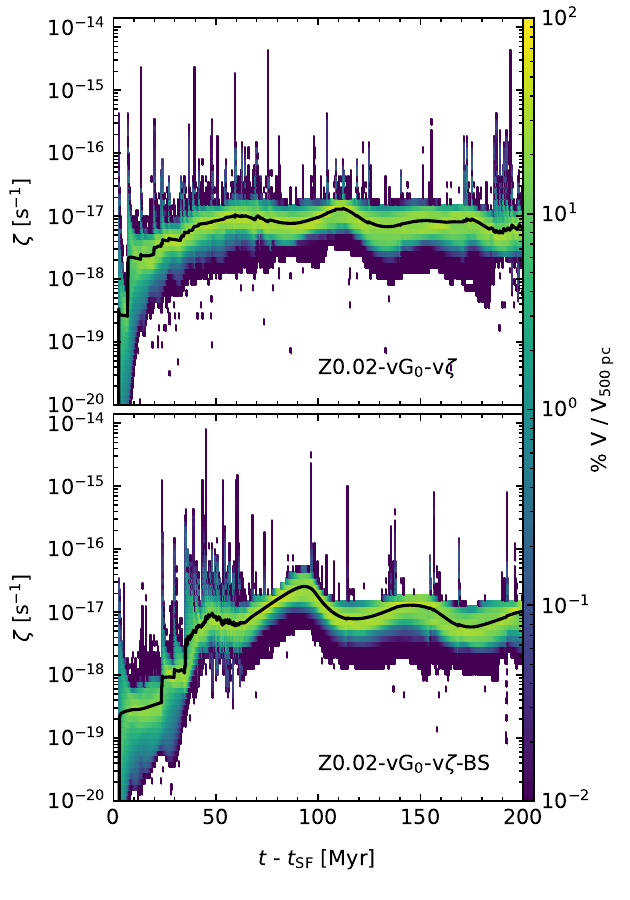}
    \caption{Local and temporal variations of $\zeta$ for runs Z0.02-vG$_0$-v$\zeta$ (top panel) and Z0.02-vG$_0$-v$\zeta$-BS (bottom panel) in a region of |$z$|~$\le$~250~pc from the disc mid-plane and in a 200 Myr interval after the onset of star formation, happening at a time $t_\mathrm{SF}$ (see Table~\ref{tab:simulations}). The colour map indicates the percentage of cells in the computational domain that, at a given time, have a certain value of $\zeta$. The black line indicates the volume-weighted mean value of $\zeta$. From this plot we can see that $\zeta$ strongly varies in both space and time.} 
    \label{fig:scatter_crir}
\end{figure}

Before analysing how the new prescription for the CR ionisation rate affects the ISM in our simulations, we briefly describe the local and temporal variation of $\zeta$, which can be seen in Fig.~\ref{fig:scatter_crir}. Here, we present $\zeta$ for the runs Z0.02-vG$_0$-v$\zeta$ (top panel) and Z0.02-vG$_0$-v$\zeta$-BS (bottom panel) in a region |$z$|~$\le$~250~pc. For every snapshot, we bin the values of $\zeta$ into 100 equally-spaced bins in logarithmic scale and we compute the percentage of cells that belong to a certain bin. By repeating the procedure for the entire time evolution of the simulation, we obtain the time-dependent histogram presented in Fig.~\ref{fig:scatter_crir}. The several peaks that can be observed correspond to the injections in CR energy subsequent to supernova explosions. In these cases, the maximum $\zeta$ reaches values up to almost 10$^{-14}$~s$^{-1}$, in accordance with the values obtained by \citealt{Pineda2024} for molecular hydrogen. The CR ionisation rate computed for molecular hydrogen is twice the value of the CR ionisation rate computed for atomic hydrogen. At times later than 50~Myr after the onset of star formation, the lower limit for $\zeta$ settles to around 10$^{-18}$~s~$^{-1}$. Moreover, the value of $\zeta$ typically varies by at least three orders of magnitude, in a range that is around one or two orders of magnitude lower than the estimates provided by \citealt{Padovani2022}. This is expected, as we have much lower star formation rates than in solar-neighbourhood conditions. Regarding the volume-weighted average of $\zeta$, we notice that around 100~Myr after the onset of star formation, which happens at a time $t_\mathrm{SF}$ (see Table~\ref{tab:simulations}), the average value settles to a value of around 10$^{-17}$~s$^{-1}$, similar to the value of the CR ionisation rate estimated by \citealt{Spitzer1968} for atomic hydrogen. Hence, the CR ionisation rate computed using this new implementation varies in space and time, affecting the local heating due to CRs and, consequently, the conditions for star formation.

\subsection{The effects of varying $\zeta$}
\label{sec:var_zeta}
\begin{figure*}
	\includegraphics[width=0.9\textwidth]{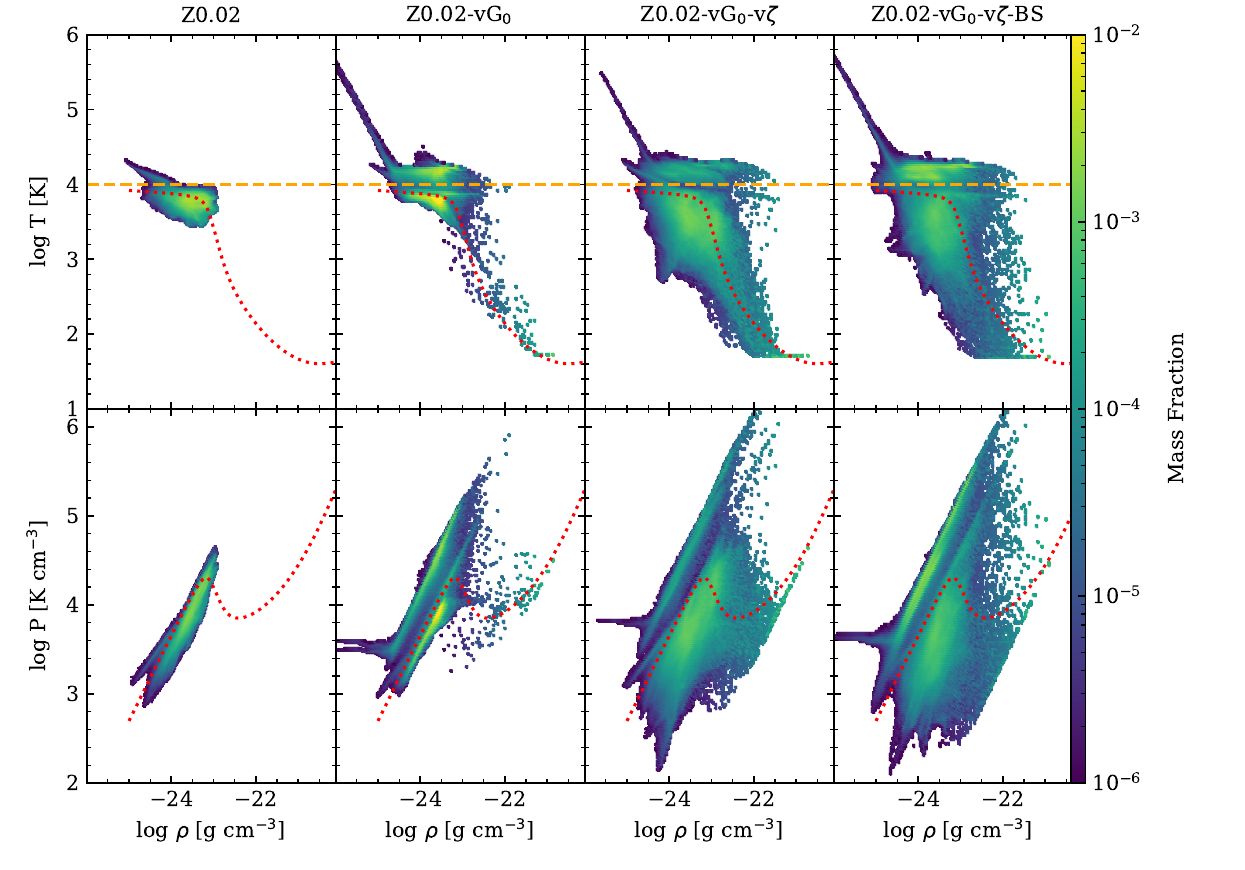}
    \caption{Temperature-density and pressure-density phase plots for the four runs, computed in a region $|z|<$~250~pc around the disc mid-plane. The red dotted line is the equilibrium curve computed for G$_0$~=~1.7, $\zeta$~=~3~$\times$~10$^{-17}$~s$^{-1}$ and a shielding column of 10$^{20}$~cm$^{-2}$. The orange dashed line indicates $T$~=~10$^4$~K. We note that the structure of the ISM changes dramatically depending on the prescription of the two heating mechanisms. Only the simulations with a variable $\zeta$ are able to cool down and form a substantial number of stars.}
    \label{fig:phase_plots}
\end{figure*}

In Fig.~\ref{fig:phase_plots} we show the temperature-density and pressure-density phase plots for the four runs. For Z0.02 the ISM is unable to cool down to lower temperatures, therefore it remains as a single-phase warm gas and star formation is inhibited. In this case, both the values of $G_0$ and $\zeta$ are such that the corresponding heating rates are too large to let the gas cool down. As these parameters are the same as the ones employed in previous works of the SILCC collaboration (e.g. \citealt{silcc6, silcc7}) for solar-metallicity conditions where star formation does take place, this shows that comparable $G_0$ and $\zeta$ values can have a different impact when the ISM is metal-poor. Considering the Z0.02-vG$_0$ simulation, which has a variable ISRF, we observe that the gas reaches lower temperatures and higher densities compared to the Z0.02 run, being able to form a few stars, however star formation is severely weakened by CR heating. A higher mass fraction of cold gas and, therefore, a larger number of stars can form in runs Z0.02-vG$_0$-v$\zeta$ and Z0.02-vG$_0$-v$\zeta$-BS, where both $G_0$ and $\zeta$ are variable. The temperature-density and pressure-density phase plots for these runs look similar as the stellar feedback changes the distribution of temperatures, pressures, and densities analogously. Therefore, it seems that a different scaling of the dust-to-gas ratio with metallicity does not significantly alter the overall thermal structure of the ISM.  

\begin{figure}
	\includegraphics[width=\columnwidth]{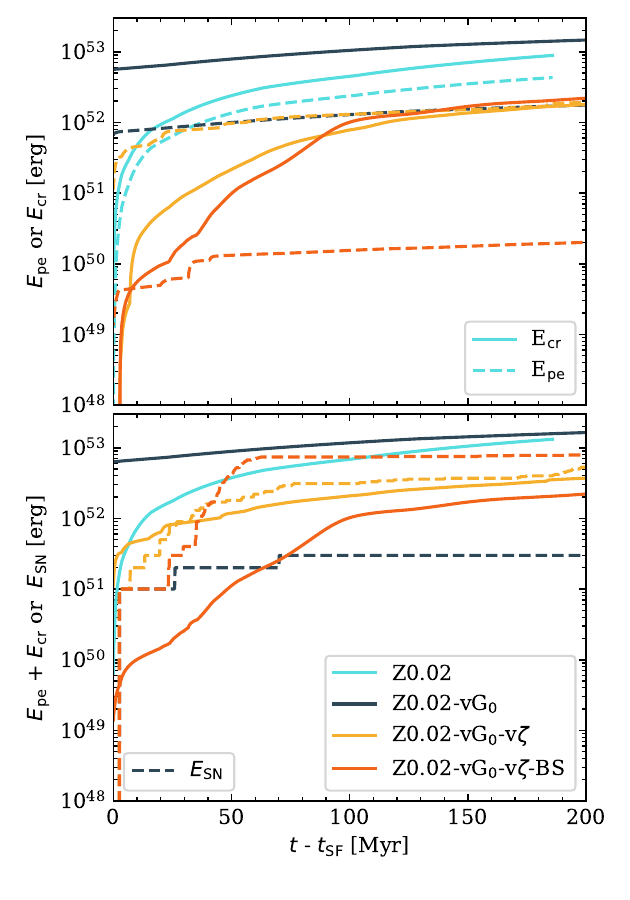}
    \caption{ \textit{Top panel}: Cumulative distributions of the energy injected in the ISM due to either PE heating or CR heating, considering a region of $\pm$~250~pc around the disc mid-plane. For the runs with constant $\zeta$, the contributions of the CR heating are higher than that of PE heating. The same behaviour is observed for run Z0.02-vG$_0$-v$\zeta$-BS. In Z0.02-vG$_0$-v$\zeta$, the opposite behaviour is observed. \textit{Bottom panel}: Cumulative distributions of the sum of the PE heating and CR heating energy injected in the ISM (solid lines), and of the energy due to supernova injections (dashed lines), in the same region as above. We note that the largest energy provided by PE and CR heating is in the case of Z0.02-vG$_0$, followed by Z0.02. For the runs with variable $\zeta$ the energy provided to the system is almost an order of magnitude lower. The total energy contribution due to both PE and CR heating is slightly lower (Z0.02-vG$_0$-v$\zeta$) or one order of magnitude lower (Z0.02-vG$_0$-v$\zeta$-BS) than the energy provided by supernovae. Note that, since the value of $t_\mathrm{SF}$ is undefined for run Z0.02, we simply show its entire evolution. }
    \label{fig:energy_CR_PE}
\end{figure}

In Fig.~\ref{fig:energy_CR_PE} we show the energy provided to the ISM by means of PE and CR heating. In the top panel, we compare the cumulative distributions of the energy due to PE heating with that of CR heating for all simulations. The energy due to CR heating is higher than that from PE heating for the runs with constant $\zeta$ (Z0.02 and Z0.02-vG$_0$) and for the Z0.02-vG$_0$-v$\zeta$-BS run. In the first case (in the Z0.02 and Z0.02-vG$_0$ runs), this is due to the fact that the CR heating for $\zeta$~=~3~$\times$~10$^{-17}$~s$^{-1}$ is higher than the PE heating rate with G$_0$~=~1.7 or less (see Fig.~\ref{fig:constant_PE_CR}). Moreover, the energy provided to the gas by CRs in the Z0.02-vG$_0$ run is higher than in Z0.02, as the few formed stars provide additional CR heating when exploding as supernovae. In the second case, in the Z0.02-vG$_0$-v$\zeta$-BS run, this is due to the power-law scaling of the dust-to-gas ratio with respect to the metallicity in run Z0.02-vG$_0$-v$\zeta$-BS, which leads to a PE heating rate that is approximately two orders of magnitude lower than in the case of a linear scaling of $d$ with metallicity. 

In the second panel of Fig.~\ref{fig:energy_CR_PE}, we show the cumulative distributions of the sum of the two contributions and we add the cumulative distributions of the energy provided by supernovae (dashed lines) for the same runs. We compute the supernova contribution by multiplying the number of supernovae by 10$^{51}$~erg.
The largest total energy due to PE and CR heating provided to the ISM are, again, found in the runs Z0.02 and Z0.02-vG$_0$. Between these, the total energy is slightly higher in the case of run Z0.02-vG$_0$ because the few generated star clusters lead to a higher PE and CR heating compared to the case where G$_0$~=~1.7. Moreover, allowing both G$_0$ and $\zeta$ to vary, decreases the total energy by around an order of magnitude compared to that of the Z0.02-vG$_0$ run. Comparing the contributions of the PE and CR heating with that due to supernovae, we note that the energy provided by supernovae is slightly higher (Z0.02-vG$_0$-v$\zeta$) or one order of magnitude higher (Z0.02-vG$_0$-v$\zeta$-BS) than the former. Only in the case of the Z0.02-vG$_0$ run the energy due to supernovae is around two orders of magnitude lower than that provided by the PE and CR heating, since this run forms only a few massive stars, and as a consequence, it has only few supernovae. 
In the Z0.02-vG$_0$-v$\zeta$-BS run the energy provided by PE and CR heating together is the lowest, and the energy provided by supernovae is the highest, compared to the other runs. The total heating, counting supernovae, PE and CR heating, is higher in the Z0.02-vG$_0$-v$\zeta$-BS than in the Z0.02-vG$_0$-v$\zeta$ run. This explains why the temperature-density phase plot for this run (see Fig.~\ref{fig:phase_plots}) shows slightly less cold gas than the corresponding plot for run Z0.02-vG$_0$-v$\zeta$.      
\begin{figure*}
	\includegraphics[width=0.9\textwidth]{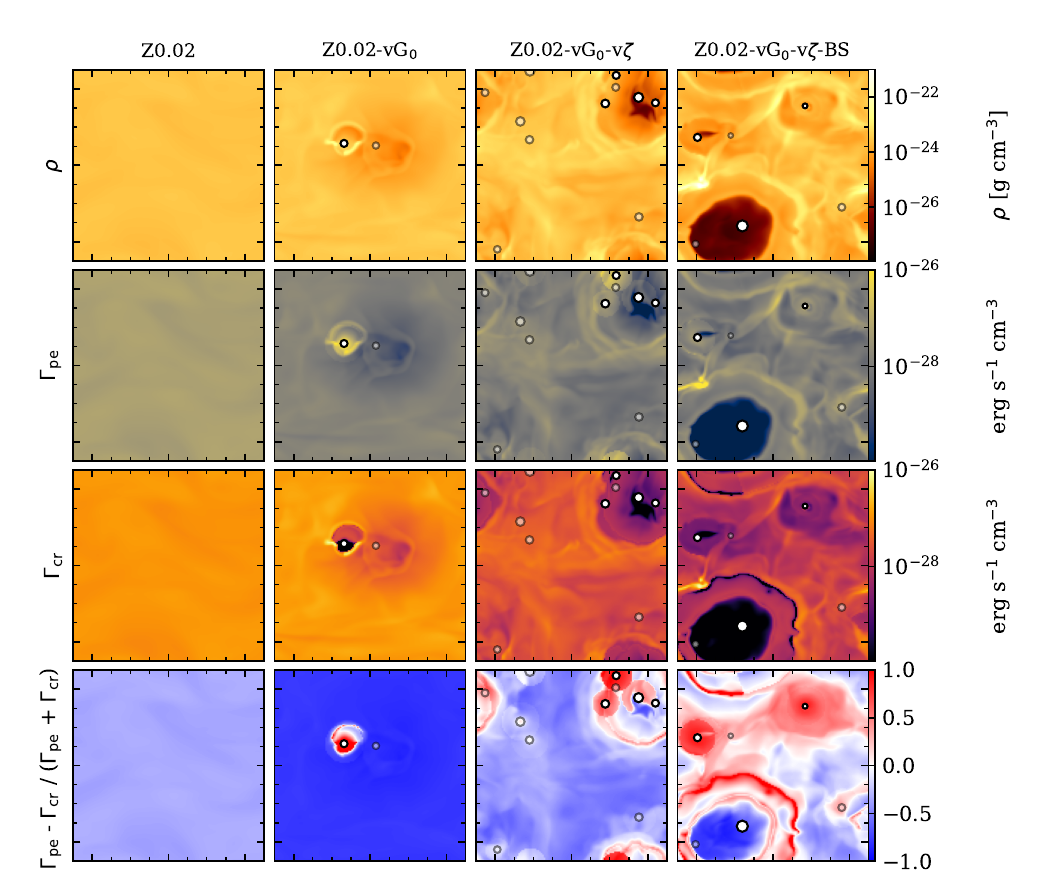}
    \caption{Slices at the disc mid-plane of the density (first row), PE heating rate (second row), CR heating rate (third row), and the difference between PE and CR heating rates normalised by the sum of the two (fourth row), for every run (columns). From left to right, the simulation time of the selected snapshots is 185.9, 144, 106 and 83.4 Myr, respectively. The last row shows that PE heating dominates near star-forming regions, whereas CR heating dominates in the more diffuse gas.}
    \label{fig:PE_CR_slice}
\end{figure*}

If both the PE and CR heating rates are allowed to vary in space and time, we can ask the question of whether there are regions where one or the other is dominant. We explore this aspect in Fig.~\ref{fig:PE_CR_slice}, where we present gas density slices taken at the mid-plane (top row), the two heating rates (second and third rows), and the normalised difference of the two (bottom row). For each run we select a snapshot where active star clusters are present near the mid-plane. In the second row, the spherical regions of high $\Gamma_\mathrm{pe}$ surrounding the star clusters are due to the $R^{-2}$ law, which is a model parameter (see Sec.~\ref{sec:AdaptiveG0}) of the G$_0$ field. For the Z0.02-vG$_0$-v$\zeta$-BS (second row, fourth column) the power-law scaling of the dust-to-gas ratio with metallicity reduces the PE heating rate by around two more orders of magnitude compared to Z0.02-vG$_0$-v$\zeta$, therefore these structures are not visible in the plot. 
In the last row we compare the PE and CR heating rates taking their difference, and normalising by their sum, such that we can obtain values in range [-1, 1] where the range (0,~1] indicates a dominating PE heating rate, and the [-1,~0) a dominating CR heating rate. We note that PE heating strongly dominates in regions close to star clusters, whereas CR heating is prevalent in the more diffuse gas further away from star-forming regions. The reason for that is the $R^{-2}$ law of the G$_0$ and the attenuation of the $\zeta$ parameter for high column densities (see Eq.~\ref{eq:cr_att}).

\subsection{Gas phases and structure}
\label{sec:gas_phases}
\begin{figure}
	\includegraphics[width=\columnwidth]{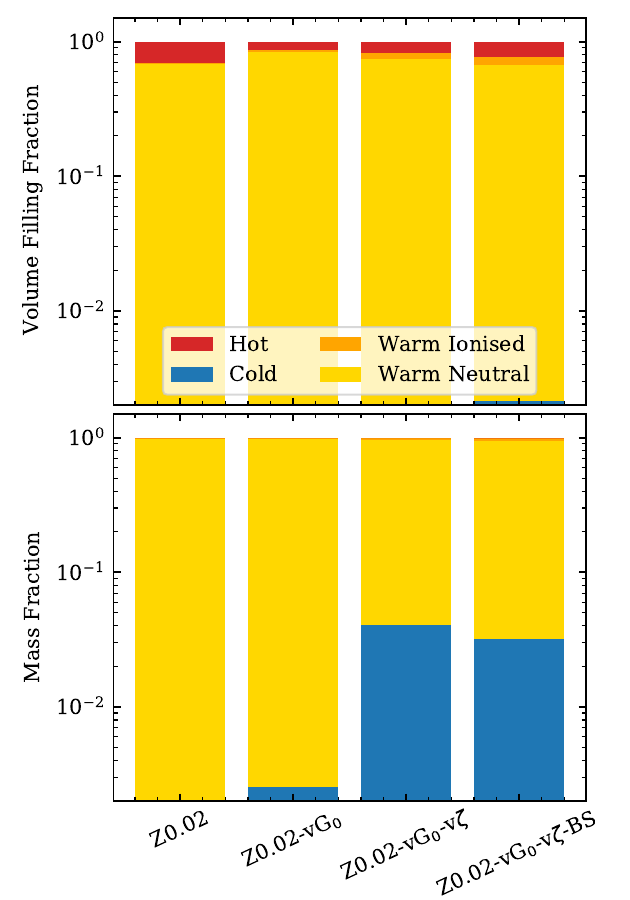}
    \caption{Average mass and volume filling fractions of the four runs, computed in a region |$z$|~$<$~250~pc around the disc mid-plane and in a time interval of 100~Myr after the onset of star formation. In run  Z0.02 no cold gas is found, whereas in run Z0.02-vG$_0$ less than 1\% of the total gas mass is cold. On the other hand, with variable $\zeta$, around 3-4\% of the mass is available as cold gas.}
    \label{fig:vff}
\end{figure}

In this section we analyse the gas phases and structure, which is important to understand star formation in low-metallicity environments. We divide the gas in a cold ($T$~$<$~300~K), a warm (300~K~$<$~$T$~$<$~3~$\times$~10$^{5}$~K) and a hot ($T$~$>$~3~$\times$~10$^{5}$~K) phase. The warm phase can either be neutral or ionised considering the neutral and ionised species present. We show the volume filling and mass fractions of these gas phases averaged over 100~Myr and computed in a region $|z|$~$<$~250~pc around the disc mid-plane in Fig.~\ref{fig:vff}. The values are summarised in Table~\ref{tab:vff}. 

In the runs with constant $\zeta$, no (Z0.02) or very little cold gas (mass fraction of 0.25\%, Z0.02-vG$_0$) forms. Therefore, the variations of the G$_0$ field do not play a key role in the gas cooling at this very low metallicity. On the other hand, in the Z0.02-vG$_0$-v$\zeta$ and Z0.02-vG$_0$-v$\zeta$-BS runs the spatial variability of $\zeta$ allows the gas to cool and form some more cold gas, whose mass fraction is around 3 -- 4\% depending on the run. As no corresponding values obtained from observations are available at this low metallicity, we compare our results with those presented in \citet{Tielens2005}, which measure 50\% of cold gas mass fraction for solar neighbourhood conditions. This value is considerably higher than that found in our runs, which can be attributed to the inefficient cooling of the gas.

\begin{table*}
	\centering
	\caption{Mass and volume filling fractions of the cold, warm neutral, warm ionised, and hot gas (see text for definition) computed for the four runs in a region of $|z|$~$<$~250~pc and averaged over 100 Myr after the onset of star formation.}
	\label{tab:vff}
	\begin{tabular}{lcccccccr} 
		\hline
		Run & $\overline{\mathrm{VFF}}_\mathrm{cold}$ & $\overline{\mathrm{VFF}}_\mathrm{WNM}$ & $\overline{\mathrm{VFF}}_\mathrm{WIM}$ & $\overline{\mathrm{VFF}}_\mathrm{hot}$ & $\overline{\mathrm{MF}}_\mathrm{cold}$ & $\overline{\mathrm{MF}}_\mathrm{WNM}$ & $\overline{\mathrm{MF}}_\mathrm{WIM}$ & $\overline{\mathrm{MF}}_\mathrm{hot}$ \\
         & $\times$~10$^{-2}$ [\%] & [\%] & [\%] & [\%] & [\%] & [\%] & [\%] & $\times$~10$^{-3}$ [\%] \\
		\hline
		Z0.02 & 0 $\pm$ 0 & 68.0 $\pm$ 0.2 & 1.6 $\pm$ 0.1 & 30.3 $\pm$ 0.2 & 0 $\pm$ 0 & 98.8 $\pm$ 0.1 & 1.14 $\pm$ 0.01 & 9.9 $\pm$ 0.1\\

        Z0.02-vG$_0$ & 0.28 $\pm$ 0.01&  84.0 $\pm$ 0.3 & 2.9 $\pm$ 0.1 & 13.1 $\pm$ 0.2 & 0.25 $\pm$ 0.01 & 98.0 $\pm$ 0.1 & 1.7 $\pm$  0.1 & 3.5 $\pm$ 0.1 \\

        Z0.02-vG$_0$-v$\zeta$ & 8.49 $\pm$ 0.17 & 74.1 $\pm$  0.6& 8.2 $\pm$ 0.2 & 17.6 $\pm$ 0.6 & 4.05 $\pm$ 0.09 & 92.1 $\pm$ 0.2 & 3.84 $\pm$ 0.14 &  11.5 $\pm$ 0.6\\
        Z0.02-vG$_0$-v$\zeta$-BS & 21.3 $\pm$ 1.1 & 67.3 $\pm$ 0.9 & 9.0 $\pm$ 0.5 & 23.5 $\pm$ 0.8 & 3.21 $\pm$ 0.15 & 91.2 $\pm$ 0.4 & 5.59 $\pm$ 0.37 & 20.6 	$\pm$ 1.0\\
		\hline
	\end{tabular}
\end{table*}

\begin{figure*}
	\includegraphics[width=0.7\textwidth]{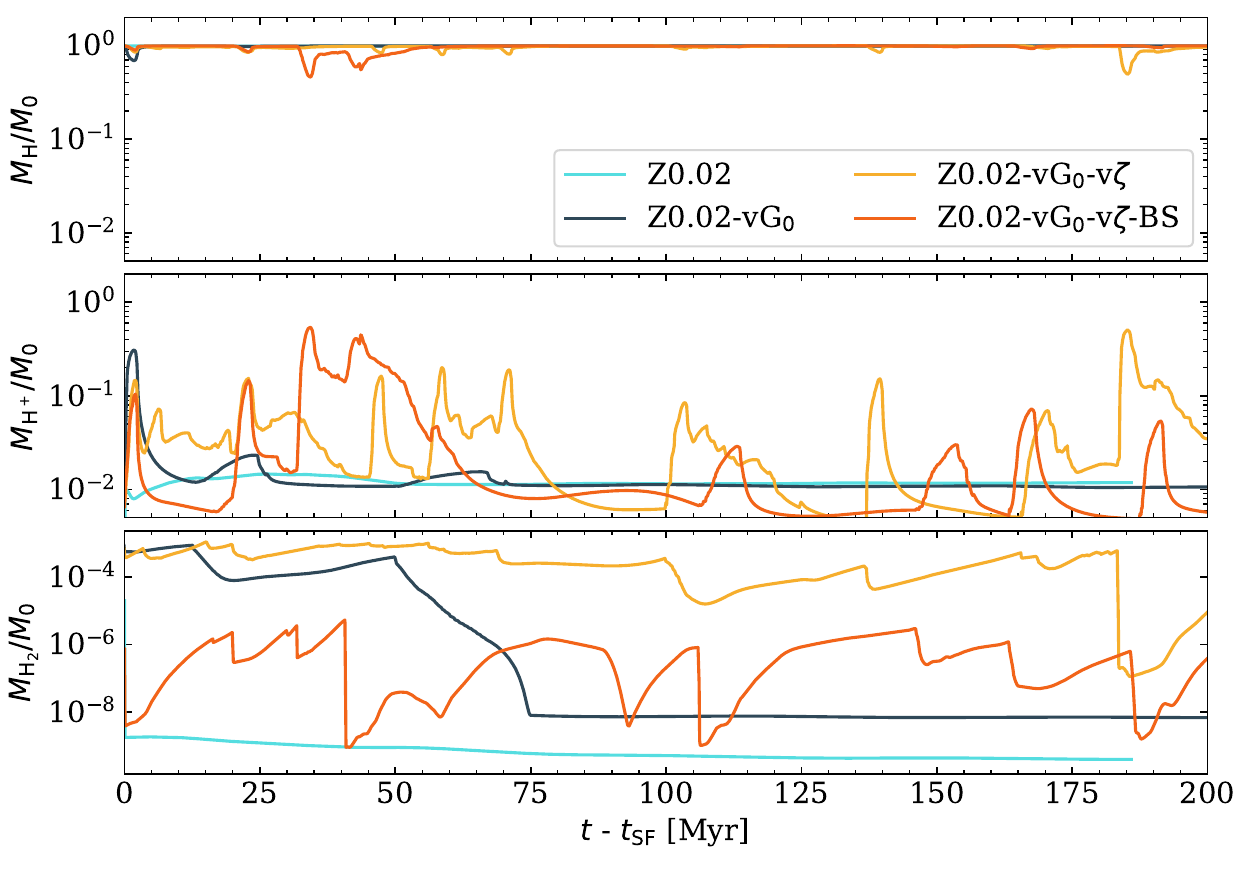}
    \caption{Volume-weighted, normalised mass fractions of H, H$^+$, H$_2$ as a function of time, computed in a region $|z|<$~250~pc. $t_\mathrm{SF}$ is the time of the first episode of star formation (see Table~\ref{tab:simulations}). \textit{Top panel}: the H mass fraction is almost always close to one, however it is slightly affected by stellar feedback for the runs that form stars. \textit{Central panel}: the ionised hydrogen fraction is affected by star formation for the runs with stars, whereas it is stable and around 10$^{-2}$ for Z0.02. \textit{Bottom panel}: 
    The H$_2$ mass fraction is highest for the Z0.02-vG$_0$-v$\zeta$ and lowest for Z0.02 (highest heating rates). Since the value of $t_\mathrm{SF}$ is undefined for the Z0.02 run, we represent here its entire evolution. }
    \label{fig:Hfractions}
\end{figure*}

In Fig.~\ref{fig:Hfractions} we show the time evolution of the atomic, ionised, and molecular hydrogen. We compute the mass in each phase within $|z|<$~250~pc divided by the total mass $M_0$ of the gas in the same region. Regarding atomic hydrogen (top panel), we notice that there is not much variation in time since, in all four simulations, most of the gas is atomic. A small change can be seen for the runs that form stars, for which a slight decrease in atomic hydrogen mass fraction corresponds to an increase in ionised hydrogen mass fraction, due to the presence of H~II regions around star clusters. However, the mass fraction of H$^+$ is almost constant and around 1\% for the Z0.02 and Z0.02-vG$_0$ runs, where no (or almost no) H~II regions are present. This value corresponds to the mass fraction of H$^+$ in the equilibrium state for a density of around 10$^{-24}$~g~cm$^{-3}$. Regarding the molecular gas (bottom panel), we note that its mass fraction is in the interval 10$^{-9}$ -- 10$^{-4}$ depending on the simulation, with the Z0.02-vG$_0$-v$\zeta$ having the highest H$_2$ mass fraction. This mass fraction range is compatible with that obtained for the G1D001 dwarf galaxy simulation in \citealt{Hu2016}, computed at a metallicity of 0.1~Z$_\odot$ and a dust-to-gas ratio of 0.01 per cent. Even though the Z0.02-vG$_0$-v$\zeta$ and Z0.02-vG$_0$-v$\zeta$-BS have almost the same amount of cold gas, there is a three to four orders of magnitude difference in the amount of molecular gas because the Z0.02-vG$_0$-v$\zeta$-BS run has a much smaller dust-to-gas ratio. For this reason, there is less formation of H$_2$ on dust grains as well as less dust shielding to protect H$_2$ from dissociation by the interstellar radiation field.

\begin{figure*}
	\includegraphics[width=0.7\textwidth]{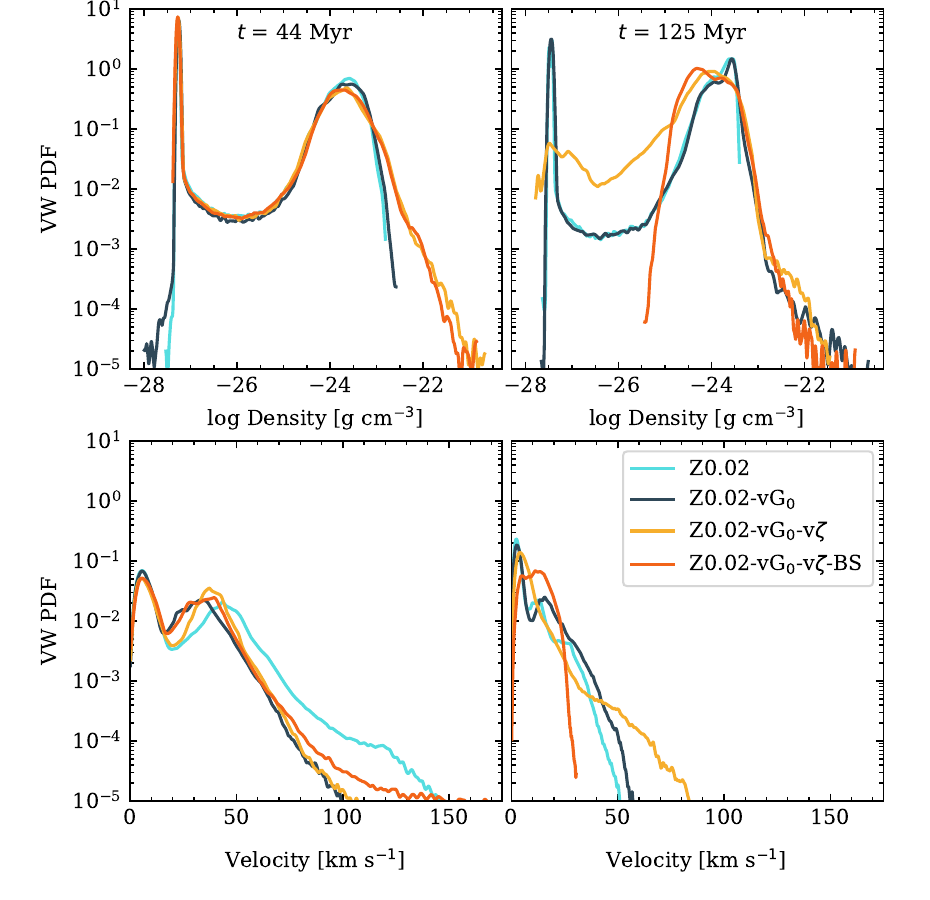}
    \caption{\textit{Top row}: Volume-weighted density PDF of the four runs right before Z0.02-vG$_0$-v$\zeta$ and Z0.02-vG$_0$-v$\zeta$ form the first star cluster (at around 44~Myr, left panel) and at a later time (125~Myr, right panel). \textit{Bottom row}: Volume-weighted velocity PDF of the four runs, computed at the same time as the corresponding above panels. Both the density and velocity PDFs are computed in a region $|z| <$~250~pc. }
    \label{fig:veldens_pdf}
\end{figure*}

We show the density (top panels) and velocity (bottom panels) probability density functions (PDFs) in Fig.~\ref{fig:veldens_pdf} at two different times, $t$~=~44~Myr and $t$~=~125~Myr, computed in a region $|z| <$~250~pc. The velocity $v$ is computed as
\begin{equation}
    v~=~\sqrt{(v_x - v_\mathrm{CM, x})^2 + (v_y - v_\mathrm{CM, y})^2 + (v_z - v_\mathrm{CM, z})^2},    
\end{equation}
where $v_i$ is the velocity of the gas in every cell and $v_\mathrm{CM, i}$ is the bulk velocity of the gas in the directions $i$~=~$x$, $y$, $z$. We choose the first time to understand the density structure of the gas right before the first star cluster is formed in the runs Z0.02-vG$_0$-v$\zeta$ and Z0.02-vG$_0$-v$\zeta$-BS and the second time to investigate the structure of the gas when their ISM has already been shaped by the stellar feedback. In the case of Z0.02-vG$_0$ the time $t$~=~125~Myr is right before the formation of the first star cluster. 

At $t$~=~44~Myr, the density PDFs of runs Z0.02-vG$_0$-v$\zeta$ and Z0.02-vG$_0$-v$\zeta$-BS exhibit a power-law tail at high densities, which could be associated with gravitational collapse \citep{Klessen2000, Slyz2005, Schneider2012, Girichidis2014}. On the other hand, no high-density regime is present at this time for runs Z0.02 and the Z0.02-vG$_0$, as the CR heating with constant $\zeta$ is preventing~/~delaying the gas cooling for run Z0.02 / Z0.02-vG$_0$. We note the presence of gas with a density lower than 10$^{-26}$~g~cm$^{-3}$, due to the fact that at this time the scaleheight of the disc is smaller than 250~pc, therefore a part of the initial ambient gas is present in this region. At $t$~=~125~Myr, however, only the two runs that have not formed stars (Z0.02 and Z0.02-vG$_0$) present this very diffuse gas. Moreover, the high-density regime is present in the case of Z0.02-vG$_0$, as the gas temperature had more time to decrease, whereas the gas in Z0.02 is simply unable to cool down and become denser.
Regarding the velocity PDFs at 44~Myr, we note one peak at around 10~km~s$^{-1}$, which is due to the $v_\mathrm{rms}$ velocity of the turbulence injected in the ISM in the first 20~Myr of evolution. We observe a second peak in the velocity range 40 -- 60~km~s$^{-1}$ due to the motion of the diffuse gas in the outer parts of the disc which is initially falling towards the disc. Toward higher velocities, the PDF decreases almost as a power law for all runs. At a time of 125~Myr, the initial turbulence has decayed, and the maximum velocities are in the range 50 -- 100~km~s$^{-1}$. These velocity values can be explained as at this time there are no active star clusters in all the runs, and in the case of the Z0.02-vG$_0$-v$\zeta$ and the Z0.02-vG$_0$-v$\zeta$-BS the gas velocity has been driven by SNe that have exploded 5-15 Myr before.

\subsection{Star formation}
\begin{figure*}
	\includegraphics[width=\textwidth]{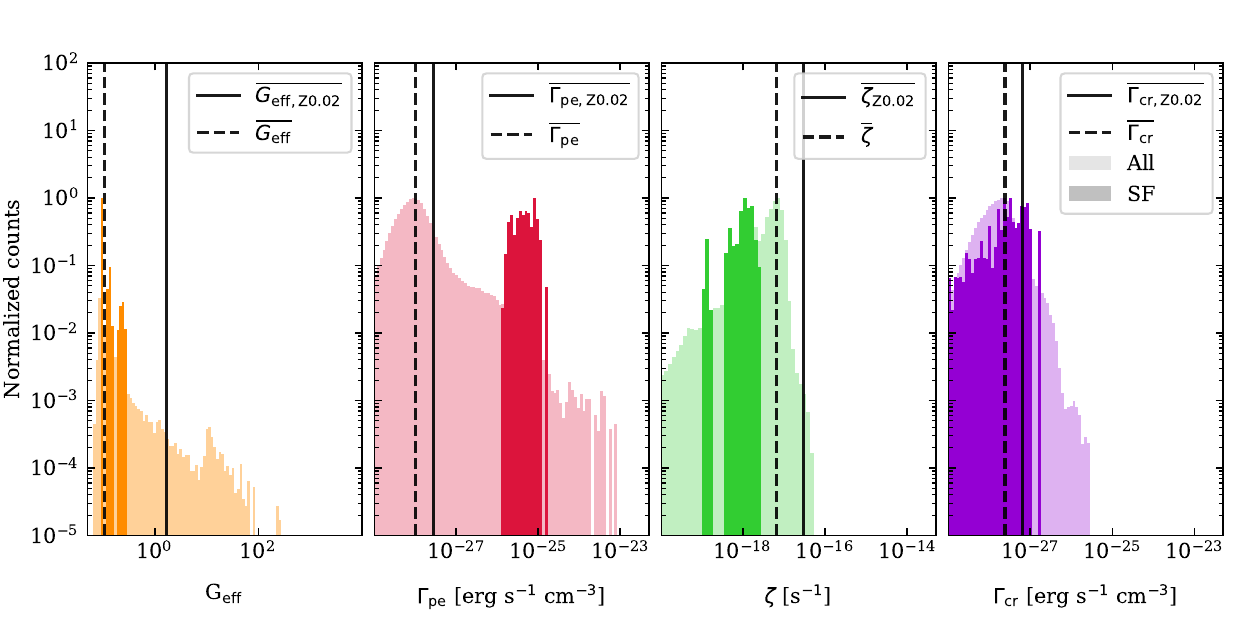}
    \caption{Initial conditions for star formation in our Z0.02-vG$_0$-v$\zeta$ simulation. For every sink particle that is created, we analyse the gas conditions in the formation region one snapshot before sink formation (see text for details).  From left to right, we show the distribution of $G_\mathrm{eff}$, $\Gamma_\mathrm{pe}$, $\zeta$, and $\Gamma_\mathrm{cr}$. Considering all star clusters, we represent the resulting distributions as dark shaded histograms. We add as transparent histograms the distributions of these quantities computed for all these snapshots and all cells in a region $|z| <$~250~pc around the mid-plane, as a comparison. All distributions are normalized to be between 0 and 1. The star-forming gas has values of $G_\mathrm{eff}$~$<$~1, $\zeta$~$\sim$~10$^{-18}$~s$^{-1}$, $\Gamma_\mathrm{pe}$~$\sim$~10$^{-26}$--10$^{-25}$~erg s$^{-1}$~cm$^{-3}$, $\Gamma_\mathrm{cr}$~$<$~10$^{-26}$~erg s$^{-1}$~cm$^{-3}$. 
    We add in the first and third panels the mean of the distributions for all gas for run Z0.02-vG$_0$-v$\zeta$ (dashed vertical lines), and, for comparison, for run Z0.02 (solid vertical lines).    }
    \label{fig:local_conditions}
\end{figure*}

In this section we analyse the properties of the gas that forms stars. As already seen above, in absence of a variable $\zeta$ the ISM is either not able to cool down (Z0.02) or, if a variable G$_0$ field is present, the gas can cool but is able to form only a few stars (Z0.02-vG$_0$) after a much longer cooling time than in case of variable $\zeta$. Therefore, unless both G$_0$ and $\zeta$ are variable quantities, no substantial star formation takes place in our simulations. In Fig.~\ref{fig:local_conditions}, we analyse four important quantities, meaning the distribution of the shielded ISRF $G_\mathrm{eff}$, the PE heating rate $\Gamma_\mathrm{pe}$, the CR ionisation rate $\zeta$, and the CR heating rate $\Gamma_\mathrm{cr}$. Taking as example the Z0.02-vG$_0$-v$\zeta$ run, we identify the coordinates where each star cluster forms, and analyse these four quantities in a snapshot exactly before the moment in which each sink particle is formed. Since the gas is moving, starting from the coordinates at which each sink particle is formed, we trace back the position that the star-forming gas would have one snapshot before the creation of the sink particle. We then consider all the cells of a region centred on these coordinates and with a radius $r_\mathrm{accr}$~$\sim$~11.7~pc, which is the accretion radius of the sink particles. We take into account only the cells of this region that have a density higher than 5~$\times$~10$^{-22}$~g~cm$^{-3}$, which is slightly lower than the density threshold $\rho_\mathrm{thr}$ used for sink formation. We represent as opaque distributions the quantities measured in these selected cells, and we refer to this gas as "star-forming". 

Additionally, we show the total distributions of the four quantities in all selected snapshots as transparent histograms. These distributions include all cells in a region $|z|$~$<$~250~pc. In order to facilitate a comparison, all distributions have been normalized such that they fall between 0 and 1. All distributions are mass-weighted. We find clear selection effects. In fact, the majority of the cells where star formation is about to take place have values of $G_\mathrm{eff} <$~1,  $\Gamma_\mathrm{pe}$ around 10$^{-26}$--10$^{-25}$ erg s$^{-1}$~cm$^{-3}$, $\zeta$~$\sim$~10$^{-18}$~s$^{-1}$, and $\Gamma_\mathrm{cr}$~$<$~10$^{-26}$ erg s$^{-1}$~cm$^{-3}$. From Eq.~\ref{eq:PE_rate} we see that $\Gamma_\mathrm{pe}$ depends on $G_\mathrm{eff}$ and the density, therefore even if the values of $G_\mathrm{eff}$ in the star-forming gas are lower than the total distribution, we still find large values of the $\Gamma_\mathrm{pe}$ because of the high density. However, the total distributions present higher values of these four quantities, meaning that photoelectric and CR heating are stronger in the cells where no star formation is taking place.  

We further analyse the impact of both, variable G$_0$ and $\zeta$, on the star formation efficiency (SFE). Therefore, in Fig.~\ref{fig:local_conditions}, we overplot the mean of $G_\mathrm{eff}$, $\Gamma_\mathrm{pe}$, $\zeta$ and $\Gamma_\mathrm{cr}$ for all gas. We show run Z0.02-vG$_0$-v$\zeta$ (i.e. $\overline{G_\mathrm{eff}}$, $\overline{\Gamma_\mathrm{pe}}$, $\overline{\zeta}$ and $\overline{\Gamma_\mathrm{cr}}$) using dashed vertical lines and run Z0.02 (i.e. $\overline{G_\mathrm{eff, Z0.02}}$, $\overline{\Gamma_\mathrm{pe, Z0.02}}$,  $\overline{\zeta_\mathrm{Z0.02}}$ and $\overline{\Gamma_\mathrm{cr, Z0.02}}$) with solid lines, respectively. 
As expected, for run Z0.02 where $G_0$ and $\zeta$ are constant, we have $\overline{G_\mathrm{eff, Z0.02}}$~=~1.7~$\times$~exp(-2.5~$A_\mathrm{V, 3D}$)~$\sim$~1.69, and $\overline{\zeta_\mathrm{Z0.02}}$~=~3~$\times$~10$^{-17}$~s$^{-1}$. The value of $\overline{G_\mathrm{eff, Z0.02}}$ does not differ much from the unattenuated value of 1.7, because assuming a minimum external column density $N_\mathrm{ext}$~=~10$^{20}$~cm$^{-2}$ and a metallicity of 0.02~Z$_\odot$, according to Eq.~\ref{eq:Av} we obtain a minimum $A_\mathrm{V, 3D}$ of around 0.001. Since the gas in run Z0.02 does not reach higher column densities than $N_\mathrm{ext}$, the minimum value of the A$_\mathrm{V, 3D}$ is similar to the average A$_\mathrm{V, 3D}$ found in this run. In run Z0.02-vG$_0$-v$\zeta$, the mean values are $\overline{G_\mathrm{eff}}$~=~0.11 and $\overline{\zeta}$~=~6.55~$\times$10$^{-18}$~s$^{-1}$. Regarding the heating rates, we find $\overline{\Gamma_\mathrm{pe}}$~=~10$^{-28}$~erg~s$^{-1}$~cm$^{-3}$, $\overline{\Gamma_\mathrm{pe, Z0.02}}$~=~2.8$\times$~10$^{-28}$, $\overline{\Gamma_\mathrm{cr}}$~=~2.4$\times$~10$^{-28}$, and $\overline{\Gamma_\mathrm{cr, Z0.02}}$~=~6.5$\times$~10$^{-28}$. 
\newline
We clearly see that choosing the constant values of G$_0$~=~1.7 and $\zeta_0$~=~3~$\times$~10$^{-17}$~s$^{-1}$ in run Z0.02 leads to higher CR and PE heating rates than the corresponding means computed for the Z0.02-vG$_0$-v$\zeta$ run. Therefore, the explanation for the inefficiency of star formation in the Z0.02 run lies in the choice of these values. At the same time, we highlight the importance of implementing variable models for G$_0$ and $\zeta$, since choosing a single value for these parameters does not reflect the diverse conditions that occur in the ISM. The wide distribution of $\zeta$, shown in the third panel of Fig.~\ref{fig:local_conditions}, confirms that a constant (lower) value of $\zeta$ is not able to properly describe the CR heating in all gas.
\newline
When looking at the star-forming gas (opaque distributions in Fig.~\ref{fig:local_conditions}), we see that star formation takes place in environments where $G_\mathrm{eff}$ is comparable to, or even higher than, the mean $\overline{G_\mathrm{eff}}$. However, $\zeta$ of the star-forming gas is actually much lower than $\overline{\zeta}$. In Sec.~\ref{sec:gas_phases}, we show that a constant $\zeta$ but variable G$_0$ (run Z0.02-vG$_0$) still leads to very little cool gas and basically no star formation. Since heating by CRs apparently dominates over PE heating at the low metallicity we investigate (see Sec.~\ref{sec:var_zeta}), choosing a constant $\zeta = \overline{\zeta}$ would again result in a lower SFE than the one we find. Hence, our results support the need for spatially and time-variable G$_0$ and $\zeta$ implementations.

\begin{figure}
	\includegraphics[width=\columnwidth]{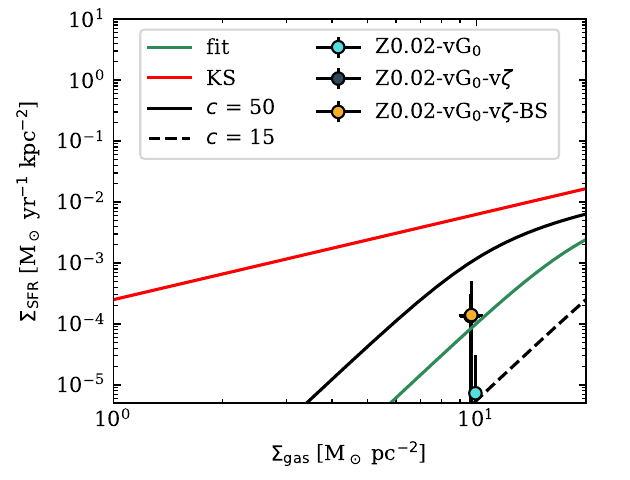}
    \caption{$\Sigma_\mathrm{SFR}$ against the gas surface density of the gas averaged over 200~Myr after the onset of star formation. We add the theoretical star formation law for low-metallicity environments from \citealt{Krumholz2009c}, computed for a value of $c$~=~50 (solid black line) and $c$~=~15 (dashed). We add the fit to our data assuming $c$ to be a free parameter (green), and the KS relation (red).The markers of the Z0.02-vG$_0$-v$\zeta$ and the Z0.02-vG$_0$-v$\zeta$-BS coincide.}
    \label{fig:ks}
\end{figure}

In Fig.~\ref{fig:ks} we represent the time-averaged star formation rate (SFR) surface density $\Sigma_\mathrm{SFR}$ as a function of $\Sigma_\mathrm{gas}$, which is the sum of the gas surface density of atomic and molecular hydrogen. The $\Sigma_\mathrm{SFR}$ is defined following \citealt{silcc3} as  
\begin{equation}
    \Sigma_\mathrm{SFR} = \frac{\mathrm{SFR}}{A} = \frac{1}{A} \sum_{i=1}^{N(t)} \frac{120 \ \mathrm{M}_\odot}{t_\mathrm{OB}},
\end{equation}
where $A$~=~(0.5~kpc)$^2$ is the area of the computational domain, $N(t)$ the number of active massive stars at time $t$, $t_\mathrm{OB}$ is the lifetime of massive stars as computed from the stellar models. Therefore, $\Sigma_\mathrm{SFR}$ corresponds to the SFR rescaled by a constant factor $A$. We average both $\Sigma_\mathrm{SFR}$ and $\Sigma_\mathrm{gas}$ in a time interval of 200~Myr after the onset of star formation in each run. As a reference we add the Kennicutt-Schmidt relation (\citealt{Kennicutt1998}, here KS), and we fit our data with the theoretical star formation law for low-metallicity environments by \citealt{Krumholz2009c} (here K09). 
For $\Sigma_\mathrm{gas} <$~85~M$_\odot$~pc$^{-2}$ (the relevant surface density range for our simulations), this relation reads
\begin{equation}
	\Sigma_\mathrm{SFR} =  f_\mathrm{H_2} \frac{\Sigma_\mathrm{gas}}{2.6 \ \mathrm{Gyr}	} \times \Big( \frac{\Sigma_\mathrm{gas}}{85 \ M_\odot \ \mathrm{pc}^{-2}} \Big) ^{-0.33},
\end{equation}
where the mass fraction of molecular gas $f_\mathrm{H_2}$ is given by
\begin{equation}
	f_\mathrm{H_2} \sim 1 - \Big[ 1  + \Big( \frac{3}{4} \frac{s}{1+\delta} \Big) ^{-5} \Big]^{-1/5},
\end{equation}
with $s$~=~$\mathrm{ln}$(1 + 0.6 $\chi$)/(0.04 $\Sigma_\mathrm{gas}$ $cZ$), $\chi$~=~0.77~(1+3.1~$Z^{0.365}$), $\delta$~=~0.0712(0.1~$s^{-1}$ + 0.675)$^{-2.8}$, with $c$ clumping factor and $Z$ the metallicity of the gas in solar units. Fitting this formula to our data gives a clumping factor of 27.3 $\pm$ 3.1, from which it follows a value of f$_\mathrm{H_2}$ $\sim$ 10$^{-2}$. This value is around one order of magnitude higher than the mass fraction of H$_2$ found for our Z0.02-vG$_0$-v$\zeta$ run. In the K09 model, $c \rightarrow 1$ if the resolution of observations is approaching 100~pc. We also add two lines computed for $c$~=~50 (solid black line) and $c$~=~15 (dashed). We note that the choice of the clumping factor strongly influences the value of the $\Sigma_\mathrm{SFR}$ computed for a given $\Sigma_\mathrm{gas}$. Additionally, we note that the average value of $\Sigma_\mathrm{SFR}$ is very similar in our Z0.02-vG$_0$-v$\zeta$ and Z0.02-vG$_0$-v$\zeta$-BS runs, despite the orders of magnitude of difference in the amount of molecular hydrogen mass present. This suggests that, in low-metallicity environments, the cold gas determines the star formation rate rather than the molecular gas \citep{Krumholz2012,Glover2012, Hu2016}. However, the $\Sigma_\mathrm{SFR}$ is around of one magnitude lower than the value predicted by the KS relation for the Z0.02-vG$_0$-v$\zeta$ and Z0.02-vG$_0$-v$\zeta$-BS runs, and two orders of magnitude lower for the Z0.02-vG$_0$ run. This discrepancy is due to the very low value of the metallicity in our runs. Measures of $\Sigma_\mathrm{SFR}$ for the IZw18 galaxy have been reported by \citealt{Aloisi1999} to be around 10$^{-2}$~M$_\odot$~yr$^{-1}$~kpc$^{-2}$ for the main body and (3 - 10)~$\times$~10$^{-3}$~M$_\odot$~yr$^{-1}$~pc$^{-2}$ for the secondary body. The $\Sigma_{\mathrm{SFR}}$ computed for the Z0.02-vG$_0$-v$\zeta$ and Z0.02-vG$_0$-v$\zeta$-BS runs is almost one order of magnitude lower than the value measured for the secondary body. This difference can be attributed to the fact that the environmental conditions of our simulations do not align perfectly with those of the IZw18 galaxy.   

\begin{figure}
	\includegraphics[width=\columnwidth]{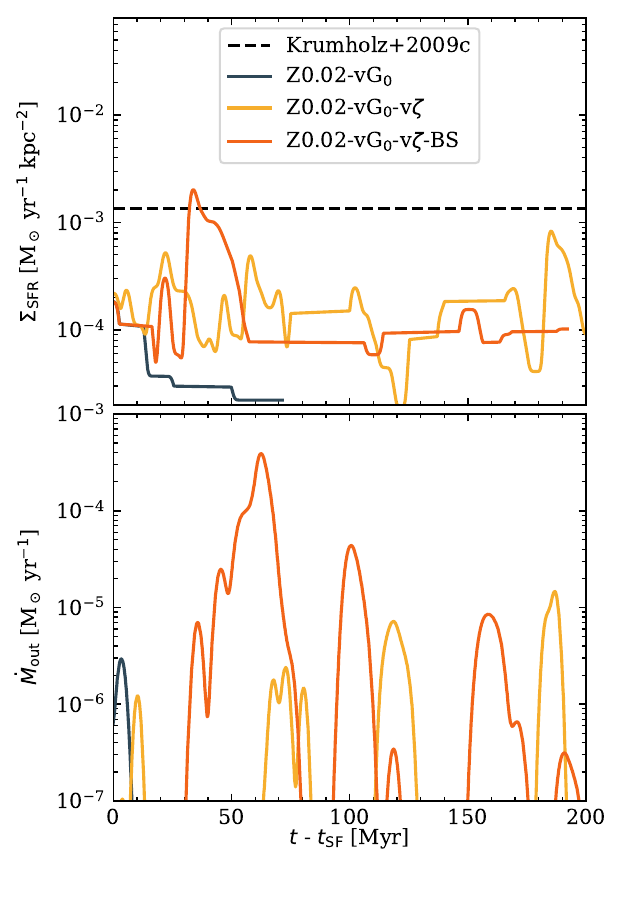}
    \caption{\textit{Top panel}: Star formation rate surface density as a function of time for the simulations that can form stars. The black line indicates the expected star formation rate surface density estimated from the \citealt{Krumholz2009c} star formation relation for a gas surface density of 10~M$_\odot$~pc$^{-2}$, at a metallicity of 0.02~Z$_\odot$ and with a clumping factor $c$~=~1. We note that this value is in accordance with the peak in $\Sigma_\mathrm{SFR}$ found for the Z0.02-vG$_0$-v$\zeta$-BS run, whereas it is higher for the other two runs.  \textit{Bottom panel}: Mass outflow rate, meaning the mass of the gas crossing a surface a $z$~=~$\pm$~1~kpc, as a function of time. }
    \label{fig:sfr_outflow}
\end{figure}

In Fig.~\ref{fig:sfr_outflow} we represent the star formation rate surface density $\Sigma_\mathrm{SFR}$ as a function of time for the runs in which stars are able to form. The star formation rate surface density oscillates in time, with peaks that reach around 2~$\times$~10$^{-3}$~M$_\odot$~yr$^{-1}$~kpc$^{-2}$ in the case of Z0.02-vG$_0$-v$\zeta$-BS. This value is around one order of magnitude less than what has been found for solar metallicity and without the variability of G$_0$ and $\zeta$ by \citealt{silcc7}. Since these setups are not directly comparable, we add a line indicating the value expected from the star formation law by \citealt{Krumholz2009c} computed for a clumping factor of one and a metallicity of 0.02~Z$_\odot$, at a gas surface density of 10~M$_\odot$~pc$^{-2}$. This value is in agreement with the peak in $\Sigma_\mathrm{SFR}$ obtained for Z0.02-vG$_0$-v$\zeta$-BS, and around one to two orders of magnitude higher for the other two runs. The reason for having so few stars can be traced back to the very little cold gas formed at this metallicity. 

The second panel of Fig.~\ref{fig:sfr_outflow} represents the mass outflow rate $\dot{M}$, meaning the mass passing through a surface at $\pm$~1~kpc from the mid-plane, in time. The peaks in $\dot{M}$ are delayed compared to the peaks in $\Sigma_\mathrm{SFR}$, as the most massive stars need at least a few million years before going off as supernovae and pushing the gas out of the disc mid-plane. However, even the largest values of $\dot{M}$ obtained in our simulations are very low. Therefore, we observe neither a thermally- nor a CR-driven outflow from the mid-plane. According to \citealt{silcc3}, at solar metallicity a hot gas volume filling fraction of at least 50\% is needed to launch an outflow from the disc. This is by far not reached in any of our runs (see Fig.~\ref{fig:vff}), as the maximum average volume filling fraction is reached by the Z0.02 run, around 30\%, followed by the Z0.02-vG$_0$-v$\zeta$-BS run, around 23\%. The amount of hot gas depends on the SN rate, which in turn depends on the star formation rate. We conclude that $\Sigma_\mathrm{SFR}$ in our low metallicity runs is too low to sustain a hot volume-filling phase and to launch a thermally driven outflow. We also do not observe a CR-driven outflow. CR injection events are rare due to the low SN rate. The CRs then diffuse rapidly, thus smoothing out local fluctuations in their energy density and pressure distributions and a CR pressure gradient that would be able to lift the gas from the disc cannot be maintained.

\begin{figure*}	\includegraphics[width=0.93\textwidth]{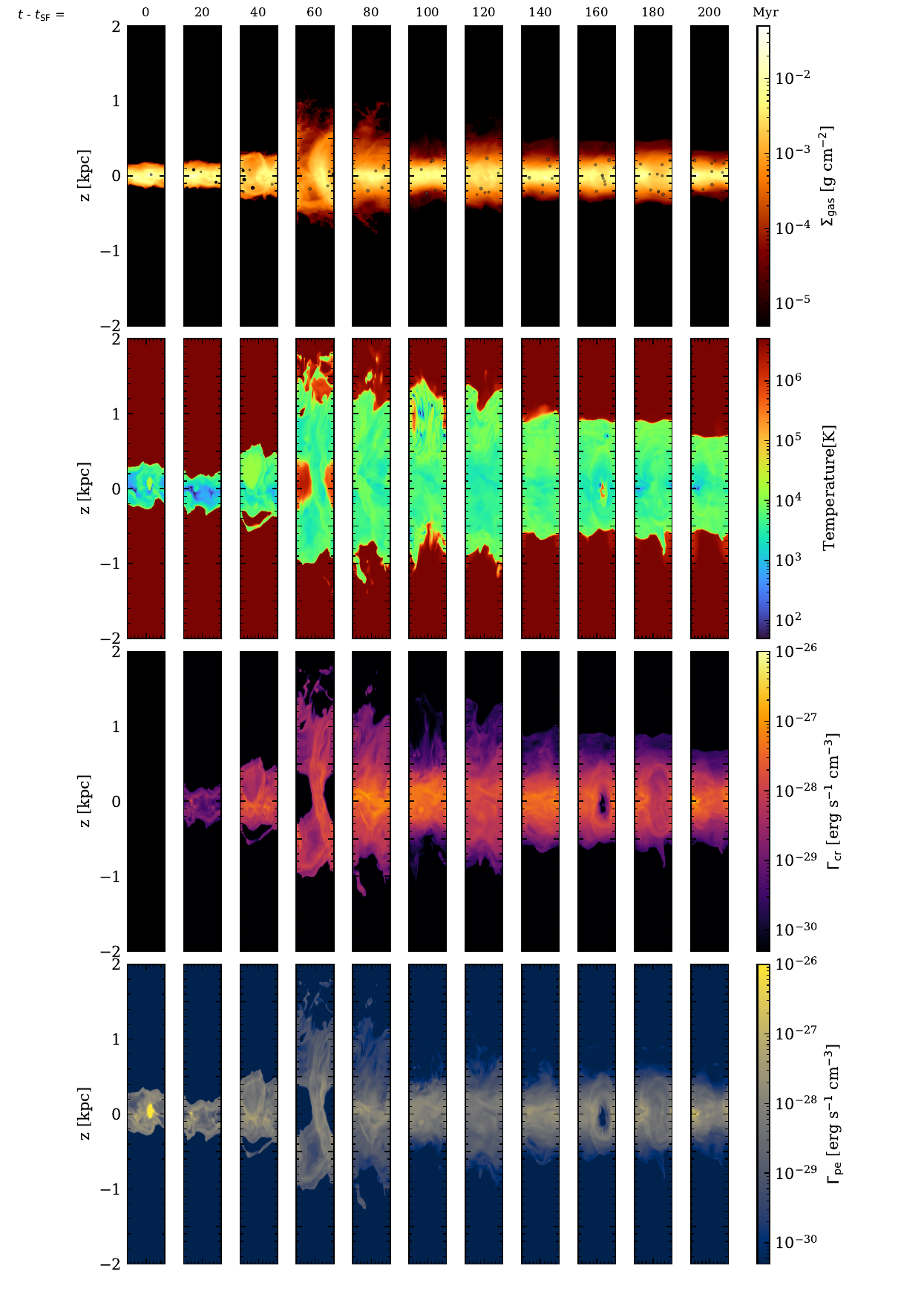}
    \caption{Time series for Z0.02-vG$_0$-v$\zeta$-BS. In the first row the projection of the density is represented, in the second the slice of temperature, in the third the slice of the CR heating rate, in the last the slice of the PE heating rate. The white circles in the first row represent the star clusters.}
    \label{fig:timeseries}
\end{figure*}

For the Z0.02-vG$_0$-v$\zeta$-BS run we show several quantities as times series in Fig.~\ref{fig:timeseries}, to represent the evolution of the gas and the related physical quantities. We show the density projection and slices for temperature, PE and CR heating. We note that no real outflow develops, as already seen in Fig.~\ref{fig:sfr_outflow}, however at around 60~Myr the gas reaches slightly less than 1~kpc in the vertical direction. After that, the evolution continues without relevant changes in the thickness of the disc. Regarding the PE and CR heating, they depend on density, therefore their distribution follows strictly that of the gas distribution.

\section{Discussion}
\label{sec:discussion}

\subsection{Computation of $\zeta$}
As already discussed above, the energy spectrum of CRs peaks at energies of a few GeV, and it declines steeply at lower and higher energies. Therefore, the energy density of CRs in the ISM is dominated by the contribution of CRs whose energy is of a few GeV. On the other hand, the CR ionisation rate is determined by low-energy CRs, meaning in the eV-MeV range. Here, we simply assume that the CR ionisation rate scales with the CR energy density, without treating the different physics that actually regulates low-energy CRs, such as energy-dependent cooling and CR transport. In this way, we assume a gray spectrum for CRs, where losses are considered to be the same regardless of the energy interval under consideration or the local thermal and magnetic conditions. However, there are six orders of magnitude of difference in energy  between the GeV CRs at the peak of the spectrum, and those mainly responsible for CR heating. Therefore, a more detailed treatment in future works is needed in order to take into account the missing physics necessary to describe the energy losses, possible re-acceleration, and diffusion of the low-energy CRs. This could for example be done by simulating live CR spectra \citep{Girichidis2020,GirichidisEtAl2022,GirichidisEtAl2024}. Our choice of using the scaling for the attenuation proposed by \citealt{Padovani2009} partly solves this issue, as the observationally-motivated relationship between $\zeta$ and the column density of the gas naturally takes into account all the relevant physics. 

\subsection{Comparison to previous works}
\citealt{girichidis2018} have implemented a variable CR ionisation rate computed by linearly scaling the energy density of CRs without any attenuation and employing a setup similar to ours. They compare their runs with a variable $\zeta$ with those where the CR ionisation rate is constant, and they observe no systematic change in the dynamics of the gas, in net contrast to our results. We can identify two reasons to explain this difference. The first lies in the different characterisation of the supernova rate. In fact, \citealt{girichidis2018} employs a fixed supernova rate, where the positions and times of each supernova are provided as a simulation input. Contrary, we follow the formation and evolution of massive stars and the resulting supernova rate depends on the population of massive stars. The local change in CR ionisation rate locally affects the CR heating, which in turn impacts the star formation, allowing us to form stars even at this very low metallicity. The second reason concerns the metallicity of the gas. \citealt{girichidis2018} study solar-neighbourhood conditions. As already seen in Sec.~\ref{sec:heating_rates}, at solar metallicity the PE heating plays a much more important role than CR heating in regulating the gas temperature. Consequently, changes in the CR ionisation rate have a much smaller impact on the phase balance between cold and warm gas in their simulations, compared to our models. This could explain why they do not observe significant variations in the dynamics of the gas when changing the local CR ionisation rate. 

\citealt{Kim2023} present the TIGRESS-NCR simulations, which adopt a simulation setup similar to that used in this work. They do not include CR transport via advection, diffusion and streaming along magnetic field lines, therefore they neglect any possible effect of the presence of CRs on the dynamics of the gas. However, they take into account CR heating employing a temporally variable CR ionisation rate which scales according to the star formation rate surface density and the gas surface density. In particular, they adopt an unattenuated value of the CR ionisation rate (in our notation) $\zeta_0$~=~2~$\times$~10$^{-16}$~s$^{-1}$~$\times$~$\Sigma '_\mathrm{SFR,40}$/$\Sigma '_\mathrm{gas}$, where $\Sigma '_\mathrm{SFR,40}$ is the star formation rate surface density computed considering the last 40~Myr of evolution divided by the solar neighbourhood value $\Sigma_\mathrm{SFR}$~=~2.5~$\times$~10$^{-3}$~M$_\odot$~kpc$^{-2}$~yr$^{-1}$, and $\Sigma '_\mathrm{gas}$ is the instantaneous gas surface density normalised by the solar-neighbourhood value $\Sigma_\mathrm{gas}$~=~10~M$_\odot$~pc$^{-2}$. For an effective shielding column density $N_\mathrm{eff}$ higher than a threshold value $N_0$~=~9.35~$\times$~10$^{20}$~cm$^{-2}$ the CR ionisation rate $\zeta_0$ is attenuated by a factor of ($N_\mathrm{eff}$/$N_0$)$^{-1}$. The exponent of their attenuation factor (-1) is assumed from \citealt{Neufeld2017}, which is more than double compared to the exponent employed in this work (-0.423), taken from \citealt{Padovani2009}. This implies that the CR ionisation rate is more heavily attenuated in the dense gas in TIGRESS-NCR simulations compared to ours meaning that, for the same $\zeta_0$, the CR heating is higher in our simulations.  Since they simulate the ISM at solar metallicity, for which different conditions apply compared to the very metal-poor medium analysed in this work, we cannot directly compare with our results. This will be done in a follow-up paper (Brugaletta et al. in prep), where we study different metallicities. However, regarding the method, we can speculate that their value of the unattenuated CR ionisation rate $\zeta_0$ would be too large to allow star formation in a metal-poor environment if $\Sigma '_\mathrm{SFR,40}$~$\sim$~$\Sigma '_\mathrm{gas}$. In fact, as seen in Fig.~\ref{fig:scatter_crir}, a value of around 10$^{-16}$~s$^{-1}$ is reached only in few cells in the vicinity of supernovae. 

An extension of this work has been presented in \citealt{Kim2024}, where they expand their analysis at metallicities from three times solar down to 0.1~Z$_\odot$, which is five times higher than the value adopted in this work. A direct comparison is, again, non trivial, and will be done in a follow-up paper (Brugaletta et al in prep.).

\subsection{Considerations for high-redshift environments}
The setup used in our simulations is suitable to describe the ISM at redshift zero, typical for nearby dwarf galaxies. However, low-metallicity environments are also common in high-redshift galaxies. To adapt our current setup in order to describe the high-redshift ISM, a few adjustments are needed. 

First of all, the temperature of the cosmic microwave background (CMB), which sets the temperature floor of the gas, depends on the redshift as
\begin{equation}
    T_\text{CMB} (z) = T_0(1 + z),
\end{equation}
where $T_\mathrm{CMB}$ is the temperature of the CMB, $z$ is the redshift, and $T_0$~=~2.725~$\pm$~0.002~K \citep{Mather1999} is the value at the present epoch. Since the temperature of the CMB depends on the redshift only linearly, it will change only slightly if we do not choose high redshifts, e.g. $z$~=~20. Therefore, the change in temperature of the CMB can probably be neglected if the redshift of interest is sufficiently low.

Moreover, H$_2$ formation would be affected by the lack of dust in pristine gas. In this work we adopt a simplified treatment for the production of H$_2$, assuming that it can form only on the surface of dust grains, following the results from \citealt{Hollenbach1989}. However, molecular hydrogen can also form in the gas, and its formation channels there can dominate over formation on dust grains at sufficiently low metallicities. For example, H$_{2}$ formation via the H$^{-}$ ion becomes more important than grain surface formation for dust-to-gas ratios $\sim 1$\% of the solar value or smaller \citep{Glover2003}, while formation via the three-body channel becomes important at high gas densities ($n>$~10$^{8}$~cm$^{-3}$, \citealt{Palla1983}) and very low metallicities 
($Z<$~10$^{-6}$~Z$_\odot$, \citealt{Omukai2005}). We have discussed above that molecular gas is not needed at this metallicity to form stars, therefore these production mechanisms do not play an important role in describing star formation at high redshift.

In addition, in galaxies with masses higher than 10$^{10}$~M$_\odot$ the gas fraction, i.e. the ratio of gas mass to stellar mass, increases as a function of redshift \citep{Carilli2013}. Moreover, according to \citealt{Markov2022}, high-redshift galaxies present a starburst phase, therefore their Kennicutt-Schmidt relation is shifted towards higher star formation rate surface densities.

\section{Summary and conclusions}
\label{sec:conclusions}

In this work we present our new treatment of the heating due to low-energy CRs implemented in the framework of the SILCC simulations. Our magneto-hydrodynamic simulations include feedback from massive stars in form of non-ionising (FUV) and ionising (EUV) radiation, stellar winds, supernovae and CRs. In our new prescription for CR heating, we treat the CR ionisation rate $\zeta$ as a variable, instead of a constant parameter, such that it varies as a function of the energy density of CRs. We assume a linear scaling of $\zeta$ with the energy density of CRs if the column density of the gas is lower than 10$^{20}$ cm$^{-2}$, otherwise we attenuate this linear scaling with a power law. The other main important heating mechanism in the ISM is photoelectric heating, which we compute from the FUV radiation field using the novel method presented in \citealt{silcc8}.

We test this implementation at a metallicity of 0.02~Z$_\odot$, for which the CR heating computed for a reference value of $\zeta$~=~3~$\times$~10$^{-17}$~s$^{-1}$ is comparable with the photoelectric heating rate computed for a reference value of G$_0$~=~1.7 and the chosen metallicity. We run four different simulations with: i) constant G$_0$ and $\zeta$, ii)  variable G$_0$ and constant $\zeta$, iii) both parameters variable and a linear scaling of the dust-to-gas ratio with the metallicity,  iv) both parameters variable and a power-law scaling of the dust-to-gas ratio with metallicity. 
Our main results are listed below. 
\begin{itemize}
    \item The variability in space and time of the CR ionisation rate plays a major role in allowing the gas to cool. In fact, having both constant G$_0$ and $\zeta$ hinders the cooling and therefore star formation. Moreover, having a variable G$_0$ but a constant $\zeta$ allows only a very small fraction of the gas to cool and form stars. Only the variability of the two parameters allows a substantial number of stars to form, however, the different scaling of the dust-to-gas ratio with metallicity does not produce important differences.
    \item The ability of the gas to cool to temperatures suitable for star formation is dictated by the amount of energy provided by photoelectric and CR heating. The runs with constant $\zeta$ are those whose total energy given by these heating mechanisms is larger, whereas the runs with a variable $\zeta$ provide almost an order of magnitude less energy. 
    \item The direct effect is the amount of cold gas formed in our runs. In fact, the runs with variable $\zeta$ result in 3 -- 4\% of the total gas mass to be cold. On the other hand, in the Z0.02-vG$_0$ run, only 0.25\% of the gas is cold. 
    \item We analyse the spatial distribution of the photoelectric and CR heating rates, and we compare them. We note that the photoelectric heating rate dominates in the immediate vicinity of star-forming regions, whereas the CR heating rate seems to dominate in the more diffuse gas, far away from massive star formation sites. 
    \item We observe different behaviours in the density PDFs, as the runs able to form stars develop a power-law tail at high density, whereas the runs with no stars do not have a high-density phase. 
    \item We note selection effects in the local distributions of $G_\mathrm{eff}$, $\zeta$, $\Gamma_\mathrm{pe}$, $\Gamma_\mathrm{cr}$ in the volume where a star cluster will be forming. For a star cluster to form, it is necessary to have a $G_\mathrm{eff} <$~1, a value of $\zeta$ around 10$^{-18}$~s$^{-1}$, $\Gamma_\mathrm{pe}$ and $\Gamma_\mathrm{cr}$ below 10$^{-25}$~erg~s$^{-1}$~cm$^{-3}$. The values of $\zeta$ for which star formation takes place are lower than the mean of the distribution computed for all gas. 
    \item Because of the above points,  substantial star formation takes place only in the runs with variable $\zeta$, where small values of $\zeta$ can be locally possible. However, the star formation rate surface density is around one order of magnitude lower than what is expected from the star formation relation for low-metallicity environments \citep{Krumholz2009c}. 
    \item Since only a few stars are formed in all runs, there are only a few supernova explosions in our simulation domain. Therefore, neither thermally- nor CR-driven outflows develop in our simulations. 
    
\end{itemize}

\section*{Acknowledgements} 
The authors thank the anonymous referee for improving this work with useful comments.VB, SW, TER, DS, and PCN thank the Deutsche Forschungsgemeinschaft (DFG) for funding through the SFB~1601 ``Habitats of massive stars across cosmic time’' (sub-projects B1, B4 and B6). SW, TER, and DS further acknowledge support by the project ''NRW-Cluster for data-intensive radio astronomy: Big Bang to Big Data (B3D)'' funded through the programme ''Profilbildung 2020'', an initiative of the Ministry of Culture and Science of the State of North Rhine-Westphalia. VB and SW thank the Bonn-Cologne Graduate School. TN acknowledges support from the DFG under Germany’s Excellence Strategy - EXC-2094 - 390783311 from the DFG Cluster of Excellence "ORIGINS". PG and SCOG acknowledge funding by the European Research Council via the
ERC Synergy Grant “ECOGAL” (project ID 855130). RW acknowledges support by the institutional project RVO:67985815.   SCOG also acknowledges support from the Heidelberg Cluster of Excellence EXC 2181 (Project-ID 390900948) `STRUCTURES: A unifying approach to emergent phenomena in the physical world, mathematics, and complex data' supported by the German Excellence Strategy. The software used in this work was in part developed
by the DOE NNSA-ASC OASCR Flash Centre at the University of Rochester \citep{Fryxell_2000, dubey2009}. Part of the data visualisation has been done with the Python package \textsc{YT} \citep{Turk2011} and the \textsc{flash\_amr\_tools} Python package (\url{https://pypi.org/project/flash-amr-tools/}) developed by PCN. The data analysis has been performed using the following Python packages: \textsc{numpy} \citep{vanderWalt2011}, \textsc{matplotlib} \citep{Hunter2007}, \textsc{h5py} \citep{Collette2020}, \textsc{IPython} \citep{Perez2007}, \textsc{SciPy} \citep{scipy}.

\section*{Data Availability}
The derived data underlying this article will be shared on reasonable request to the corresponding author. The simulation data will be made available on the SILCC data web page: \url{http://silcc.mpa-garching.mpg.de}.



\bibliographystyle{mnras}
\bibliography{main} 

\begin{thebibliography}{}
\makeatletter
\relax
\def\mn@urlcharsother{\let\do\@makeother \do\$\do\&\do\#\do\^\do\_\do\%\do\~}
\def\mn@doi{\begingroup\mn@urlcharsother \@ifnextchar [ {\mn@doi@} {\mn@doi@[]}}
\def\mn@doi@[#1]#2{\def\@tempa{#1}\ifx\@tempa\@empty \href {http://dx.doi.org/#2} {doi:#2}\else \href {http://dx.doi.org/#2} {#1}\fi \endgroup}
\def\mn@eprint#1#2{\mn@eprint@#1:#2::\@nil}
\def\mn@eprint@arXiv#1{\href {http://arxiv.org/abs/#1} {{\tt arXiv:#1}}}
\def\mn@eprint@dblp#1{\href {http://dblp.uni-trier.de/rec/bibtex/#1.xml} {dblp:#1}}
\def\mn@eprint@#1:#2:#3:#4\@nil{\def\@tempa {#1}\def\@tempb {#2}\def\@tempc {#3}\ifx \@tempc \@empty \let \@tempc \@tempb \let \@tempb \@tempa \fi \ifx \@tempb \@empty \def\@tempb {arXiv}\fi \@ifundefined {mn@eprint@\@tempb}{\@tempb:\@tempc}{\expandafter \expandafter \csname mn@eprint@\@tempb\endcsname \expandafter{\@tempc}}}

\bibitem[\protect\citeauthoryear{{Ackermann} et~al.,}{{Ackermann} et~al.}{2013}]{Ackermann2013}
{Ackermann} M.,  et~al., 2013, \mn@doi [Science] {10.1126/science.1231160}, \href {https://ui.adsabs.harvard.edu/abs/2013Sci...339..807A/abstract} {339, 807}

\bibitem[\protect\citeauthoryear{{Aloisi}, {Tosi}  \& {Greggio}}{{Aloisi} et~al.}{1999}]{Aloisi1999}
{Aloisi} A.,  {Tosi} M.,   {Greggio} L.,  1999, \mn@doi [\aj] {10.1086/300924}, \href {https://ui.adsabs.harvard.edu/abs/1999AJ....118..302A} {118, 302}

\bibitem[\protect\citeauthoryear{{Axford}, {Leer}  \& {et al.,}}{{Axford} et~al.}{1978}]{Axford1978}
{Axford} W.~I.,  {Leer} E.,   {et al.,} 1978, in {Dergachev} V.~A.,  {Kocharov} G.~E.,  eds, Cosmophysics. pp 125--134

\bibitem[\protect\citeauthoryear{{Baade} \& {Zwicky}}{{Baade} \& {Zwicky}}{1934}]{Baade1934}
{Baade} W.,  {Zwicky} F.,  1934, \mn@doi [Physical Review] {10.1103/PhysRev.46.76.2}, \href {https://ui.adsabs.harvard.edu/abs/1934PhRv...46...76B} {46, 76}

\bibitem[\protect\citeauthoryear{{Bakes} \& {Tielens}}{{Bakes} \& {Tielens}}{1994}]{Bakes1994}
{Bakes} E.~L.~O.,  {Tielens} A.~G.~G.~M.,  1994, \mn@doi [\apj] {10.1086/174188}, \href {https://ui.adsabs.harvard.edu/abs/1994ApJ...427..822B} {427, 822}

\bibitem[\protect\citeauthoryear{Bate, Bonnell  \& Price}{Bate et~al.}{1995}]{bate1995}
Bate M.~R.,  Bonnell I.~A.,   Price N.~M.,  1995, \mn@doi [MNRAS] {10.1093/mnras/277.2.362}, \href {https://ui.adsabs.harvard.edu/abs/1995MNRAS.277..362B/abstract} {277, 362}

\bibitem[\protect\citeauthoryear{{Bell}}{{Bell}}{1978a}]{Bell1978a}
{Bell} A.~R.,  1978a, \mn@doi [\mnras] {10.1093/mnras/182.2.147}, \href {https://ui.adsabs.harvard.edu/abs/1978MNRAS.182..147B} {182, 147}

\bibitem[\protect\citeauthoryear{{Bell}}{{Bell}}{1978b}]{Bell1978b}
{Bell} A.~R.,  1978b, \mn@doi [\mnras] {10.1093/mnras/182.3.443}, \href {https://ui.adsabs.harvard.edu/abs/1978MNRAS.182..443B} {182, 443}

\bibitem[\protect\citeauthoryear{{Bergin} \& {Tafalla}}{{Bergin} \& {Tafalla}}{2007}]{Bergin2007}
{Bergin} E.~A.,  {Tafalla} M.,  2007, \mn@doi [\araa] {10.1146/annurev.astro.45.071206.100404}, \href {https://ui.adsabs.harvard.edu/abs/2007ARA&A..45..339B} {45, 339}

\bibitem[\protect\citeauthoryear{{Bergin}, {Hartmann}  \& {et al.,}}{{Bergin} et~al.}{2004}]{Bergin2004}
{Bergin} E.~A.,  {Hartmann} L.~W.,   {et al.,} 2004, \mn@doi [\apj] {10.1086/422578}, \href {https://ui.adsabs.harvard.edu/abs/2004ApJ...612..921B} {612, 921}

\bibitem[\protect\citeauthoryear{{Bialy} \& {Sternberg}}{{Bialy} \& {Sternberg}}{2019}]{Bialy2019}
{Bialy} S.,  {Sternberg} A.,  2019, \mn@doi [\apj] {10.3847/1538-4357/ab2fd1}, \href {https://ui.adsabs.harvard.edu/abs/2019ApJ...881..160B} {881, 160}

\bibitem[\protect\citeauthoryear{{Blandford} \& {Ostriker}}{{Blandford} \& {Ostriker}}{1978}]{BlandfordOstriker1978}
{Blandford} R.~D.,  {Ostriker} J.~P.,  1978, \mn@doi [\apjl] {10.1086/182658}, \href {https://ui.adsabs.harvard.edu/abs/1978ApJ...221L..29B} {221, L29}

\bibitem[\protect\citeauthoryear{{Bohlin}, {Savage}  \& {Drake}}{{Bohlin} et~al.}{1978}]{Bohlin1978}
{Bohlin} R.~C.,  {Savage} B.~D.,   {Drake} J.~F.,  1978, \mn@doi [\apj] {10.1086/156357}, \href {https://ui.adsabs.harvard.edu/abs/1978ApJ...224..132B} {224, 132}

\bibitem[\protect\citeauthoryear{{Boulares} \& {Cox}}{{Boulares} \& {Cox}}{1990}]{Boulares1990}
{Boulares} A.,  {Cox} D.~P.,  1990, \mn@doi [\apj] {10.1086/169509}, \href {https://ui.adsabs.harvard.edu/abs/1990ApJ...365..544B} {365, 544}

\bibitem[\protect\citeauthoryear{{Brott} et~al.,}{{Brott} et~al.}{2011}]{brott2011}
{Brott} I.,  et~al., 2011, \mn@doi [\aap] {10.1051/0004-6361/201016113}, \href {https://ui.adsabs.harvard.edu/abs/2011A%26A...530A.115B/abstract} {530, A115}

\bibitem[\protect\citeauthoryear{{Cabedo}, {Maury}  \& {et al., }}{{Cabedo} et~al.}{2023}]{Cabedo2023}
{Cabedo} V.,  {Maury} A.,   {et al., } 2023, \mn@doi [\aap] {10.1051/0004-6361/202243813}, \href {https://ui.adsabs.harvard.edu/abs/2023A&A...669A..90C} {669, A90}

\bibitem[\protect\citeauthoryear{{Carilli} \& {Walter}}{{Carilli} \& {Walter}}{2013}]{Carilli2013}
{Carilli} C.~L.,  {Walter} F.,  2013, \mn@doi [\araa] {10.1146/annurev-astro-082812-140953}, \href {https://ui.adsabs.harvard.edu/abs/2013ARA&A..51..105C} {51, 105}

\bibitem[\protect\citeauthoryear{{Caselli}, {Walmsley}, {Terzieva}  \& {Herbst}}{{Caselli} et~al.}{1998}]{Caselli1998}
{Caselli} P.,  {Walmsley} C.~M.,  {Terzieva} R.,   {Herbst} E.,  1998, \mn@doi [\apj] {10.1086/305624}, \href {https://ui.adsabs.harvard.edu/abs/1998ApJ...499..234C} {499, 234}

\bibitem[\protect\citeauthoryear{{Ceccarelli}, {Dominik}  \& {et al.,}}{{Ceccarelli} et~al.}{2014}]{Ceccarelli2014}
{Ceccarelli} C.,  {Dominik} C.,   {et al.,} 2014, \mn@doi [\apjl] {10.1088/2041-8205/790/1/L1}, \href {https://ui.adsabs.harvard.edu/abs/2014ApJ...790L...1C} {790, L1}

\bibitem[\protect\citeauthoryear{{Collette} et~al.,}{{Collette} et~al.}{2020}]{Collette2020}
{Collette} A.,  et~al., 2020, h5py/h5py: 3.1.0, \mn@doi{10.5281/zenodo.4250762}

\bibitem[\protect\citeauthoryear{{Cox}}{{Cox}}{2005}]{Cox2005}
{Cox} D.~P.,  2005, \mn@doi [\araa] {10.1146/annurev.astro.43.072103.150615}, \href {https://ui.adsabs.harvard.edu/abs/2005ARA&A..43..337C} {43, 337}

\bibitem[\protect\citeauthoryear{{Cummings} et~al.,}{{Cummings} et~al.}{2016}]{Cummings2016}
{Cummings} A.~C.,  et~al., 2016, \mn@doi [\apj] {10.3847/0004-637X/831/1/18}, \href {https://ui.adsabs.harvard.edu/abs/2016ApJ...831...18C} {831, 18}

\bibitem[\protect\citeauthoryear{{Curti} et~al.,}{{Curti} et~al.}{2024}]{Curti2024}
{Curti} M.,  et~al., 2024, \mn@doi [\aap] {10.1051/0004-6361/202346698}, \href {https://ui.adsabs.harvard.edu/abs/2024A&A...684A..75C} {684, A75}

\bibitem[\protect\citeauthoryear{{Dinnbier} \& {Walch}}{{Dinnbier} \& {Walch}}{2020}]{Dinnbier2020}
{Dinnbier} F.,  {Walch} S.,  2020, \mn@doi [\mnras] {10.1093/mnras/staa2560}, \href {https://ui.adsabs.harvard.edu/abs/2020MNRAS.499..748D/abstract} {499, 748}

\bibitem[\protect\citeauthoryear{{Draine}}{{Draine}}{1978}]{draine1978}
{Draine} B.~T.,  1978, \mn@doi [\apjs] {10.1086/190513}, 36, 595

\bibitem[\protect\citeauthoryear{{Draine} \& {Bertoldi}}{{Draine} \& {Bertoldi}}{1996}]{DraineBertoldi1996}
{Draine} B.~T.,  {Bertoldi} F.,  1996, \mn@doi [\apj] {10.1086/177689}, \href {https://ui.adsabs.harvard.edu/abs/1996ApJ...468..269D} {468, 269}

\bibitem[\protect\citeauthoryear{{Draine} et~al.,}{{Draine} et~al.}{2007}]{draine2007}
{Draine} B.~T.,  et~al., 2007, \mn@doi [\apj] {10.1086/518306}, \href {https://ui.adsabs.harvard.edu/abs/2007ApJ...663..866D} {663, 866}

\bibitem[\protect\citeauthoryear{Dubey, Reid  \& Fisher}{Dubey et~al.}{2008}]{Dubey_2008}
Dubey A.,  Reid L.~B.,   Fisher R.,  2008, \mn@doi [Physica Scripta] {10.1088/0031-8949/2008/t132/014046}, \href {https://ui.adsabs.harvard.edu/abs/2008PhST..132a4046D/abstract} {T132, 014046}

\bibitem[\protect\citeauthoryear{{Dubey}, {Reid}  \& {et al.,}}{{Dubey} et~al.}{2009}]{dubey2009}
{Dubey} A.,  {Reid} L.~B.,   {et al.,} 2009, \mn@doi [arXiv e-prints] {10.48550/arXiv.0903.4875}, \href {https://ui.adsabs.harvard.edu/abs/2009arXiv0903.4875D} {p. arXiv:0903.4875}

\bibitem[\protect\citeauthoryear{{Favre} et~al.,}{{Favre} et~al.}{2018}]{Favre2018}
{Favre} C.,  et~al., 2018, \mn@doi [\apj] {10.3847/1538-4357/aabfd4}, \href {https://ui.adsabs.harvard.edu/abs/2018ApJ...859..136F} {859, 136}

\bibitem[\protect\citeauthoryear{Federrath, Banerjee, Clark  \& Klessen}{Federrath et~al.}{2010}]{Federrath_2010}
Federrath C.,  Banerjee R.,  Clark P.~C.,   Klessen R.~S.,  2010, \mn@doi [ApJ] {10.1088/0004-637x/713/1/269}, \href {https://ui.adsabs.harvard.edu/abs/2010ApJ...713..269F/abstract} {713, 269}

\bibitem[\protect\citeauthoryear{{Field}, {Goldsmith}  \& {Habing}}{{Field} et~al.}{1969}]{Field1969}
{Field} G.~B.,  {Goldsmith} D.~W.,   {Habing} H.~J.,  1969, \mn@doi [\apjl] {10.1086/180324}, \href {https://ui.adsabs.harvard.edu/abs/1969ApJ...155L.149F} {155, L149}

\bibitem[\protect\citeauthoryear{{Fontani} et~al.,}{{Fontani} et~al.}{2017}]{Fontani2017}
{Fontani} F.,  et~al., 2017, \mn@doi [\aap] {10.1051/0004-6361/201730527}, \href {https://ui.adsabs.harvard.edu/abs/2017A&A...605A..57F} {605, A57}

\bibitem[\protect\citeauthoryear{{French}}{{French}}{1980}]{French1980}
{French} H.~B.,  1980, \mn@doi [\apj] {10.1086/158205}, \href {https://ui.adsabs.harvard.edu/abs/1980ApJ...240...41F} {240, 41}

\bibitem[\protect\citeauthoryear{Fryxell et~al.,}{Fryxell et~al.}{2000}]{Fryxell_2000}
Fryxell B.,  et~al., 2000, \mn@doi [ApJS] {10.1086/317361}, \href {https://ui.adsabs.harvard.edu/abs/2000ApJS..131..273F/abstract} {131, 273}

\bibitem[\protect\citeauthoryear{{Gatto} et~al.,}{{Gatto} et~al.}{2015}]{gatto2015}
{Gatto} A.,  et~al., 2015, \mn@doi [\mnras] {10.1093/mnras/stv324}, \href {https://ui.adsabs.harvard.edu/abs/2015MNRAS.449.1057G/abstract} {449, 1057}

\bibitem[\protect\citeauthoryear{{Gatto} et~al.,}{{Gatto} et~al.}{2017}]{silcc3}
{Gatto} A.,  et~al., 2017, \mn@doi [\mnras] {10.1093/mnras/stw3209}, \href {https://ui.adsabs.harvard.edu/abs/2017MNRAS.466.1903G/abstract} {466, 1903}

\bibitem[\protect\citeauthoryear{{Ginzburg} \& {Syrovatskii}}{{Ginzburg} \& {Syrovatskii}}{1964}]{Ginzburg1964}
{Ginzburg} V.~L.,  {Syrovatskii} S.~I.,  1964, {The Origin of Cosmic Rays}

\bibitem[\protect\citeauthoryear{{Girichidis}, {Konstandin}  \& {et al., }}{{Girichidis} et~al.}{2014}]{Girichidis2014}
{Girichidis} P.,  {Konstandin} L.,   {et al., } 2014, \mn@doi [\apj] {10.1088/0004-637X/781/2/91}, \href {https://ui.adsabs.harvard.edu/abs/2014ApJ...781...91G} {781, 91}

\bibitem[\protect\citeauthoryear{{Girichidis} et~al.,}{{Girichidis} et~al.}{2016a}]{silcc2}
{Girichidis} P.,  et~al., 2016a, \mn@doi [\mnras] {10.1093/mnras/stv2742}, \href {https://ui.adsabs.harvard.edu/abs/2016MNRAS.456.3432G/abstract} {456, 3432}

\bibitem[\protect\citeauthoryear{{Girichidis} et~al.,}{{Girichidis} et~al.}{2016b}]{girichidis2016}
{Girichidis} P.,  et~al., 2016b, \mn@doi [\apjl] {10.3847/2041-8205/816/2/L19}, \href {https://ui.adsabs.harvard.edu/abs/2016ApJ...816L..19G} {816, L19}

\bibitem[\protect\citeauthoryear{{Girichidis}, {Naab}, {Hanasz}  \& {Walch}}{{Girichidis} et~al.}{2018a}]{girichidis2018}
{Girichidis} P.,  {Naab} T.,  {Hanasz} M.,   {Walch} S.,  2018a, \mn@doi [\mnras] {10.1093/mnras/sty1653}, \href {https://ui.adsabs.harvard.edu/abs/2018MNRAS.479.3042G} {479, 3042}

\bibitem[\protect\citeauthoryear{{Girichidis}, {Seifried}  \& {et al., }}{{Girichidis} et~al.}{2018b}]{silcc5}
{Girichidis} P.,  {Seifried} D.,   {et al., } 2018b, \mn@doi [\mnras] {10.1093/mnras/sty2016}, \href {https://ui.adsabs.harvard.edu/abs/2018MNRAS.480.3511G/abstract} {480, 3511}

\bibitem[\protect\citeauthoryear{{Girichidis}, {Pfrommer}, {Hanasz}  \& {Naab}}{{Girichidis} et~al.}{2020}]{Girichidis2020}
{Girichidis} P.,  {Pfrommer} C.,  {Hanasz} M.,   {Naab} T.,  2020, \mn@doi [\mnras] {10.1093/mnras/stz2961}, \href {https://ui.adsabs.harvard.edu/abs/2020MNRAS.491..993G} {491, 993}

\bibitem[\protect\citeauthoryear{{Girichidis}, {Pfrommer}  \& {et al., }}{{Girichidis} et~al.}{2022}]{GirichidisEtAl2022}
{Girichidis} P.,  {Pfrommer} C.,   {et al., } 2022, \mn@doi [\mnras] {10.1093/mnras/stab3462}, \href {https://ui.adsabs.harvard.edu/abs/2022MNRAS.510.3917G} {510, 3917}

\bibitem[\protect\citeauthoryear{{Girichidis}, {Werhahn}  \& {et al., }}{{Girichidis} et~al.}{2024}]{GirichidisEtAl2024}
{Girichidis} P.,  {Werhahn} M.,   {et al., } 2024, \mn@doi [\mnras] {10.1093/mnras/stad3628}, \href {https://ui.adsabs.harvard.edu/abs/2024MNRAS.527.10897} {527, 10897}

\bibitem[\protect\citeauthoryear{{Glassgold} \& {Langer}}{{Glassgold} \& {Langer}}{1974}]{Glassgold1974}
{Glassgold} A.~E.,  {Langer} W.~D.,  1974, \mn@doi [\apj] {10.1086/153130}, \href {https://ui.adsabs.harvard.edu/abs/1974ApJ...193...73G} {193, 73}

\bibitem[\protect\citeauthoryear{{Glover}}{{Glover}}{2003}]{Glover2003}
{Glover} S. C.~O.,  2003, \mn@doi [\apj] {10.1086/345684}, \href {https://ui.adsabs.harvard.edu/abs/2003ApJ...584..331G} {584, 331}

\bibitem[\protect\citeauthoryear{{Glover} \& {Clark}}{{Glover} \& {Clark}}{2012}]{Glover2012}
{Glover} S. C.~O.,  {Clark} P.~C.,  2012, \mn@doi [\mnras] {10.1111/j.1365-2966.2011.19648.x}, \href {https://ui.adsabs.harvard.edu/abs/2012MNRAS.421....9G} {421, 9}

\bibitem[\protect\citeauthoryear{{Glover} \& {Mac Low}}{{Glover} \& {Mac Low}}{2007a}]{glover2007a}
{Glover} S. C.~O.,  {Mac Low} M.-M.,  2007a, \mn@doi [\apjs] {10.1086/512238}, \href {https://ui.adsabs.harvard.edu/abs/2007ApJS..169..239G/abstract} {169, 239}

\bibitem[\protect\citeauthoryear{{Glover} \& {Mac Low}}{{Glover} \& {Mac Low}}{2007b}]{Glover_2007b}
{Glover} S. C.~O.,  {Mac Low} M.-M.,  2007b, \mn@doi [ApJ] {10.1086/512227}, \href {https://ui.adsabs.harvard.edu/abs/2007ApJS..169..239G/abstract} {659, 1317}

\bibitem[\protect\citeauthoryear{{Gnat} \& {Ferland}}{{Gnat} \& {Ferland}}{2012}]{Gnat2012}
{Gnat} O.,  {Ferland} G.~J.,  2012, \mn@doi [\apjs] {10.1088/0067-0049/199/1/20}, \href {https://ui.adsabs.harvard.edu/abs/2012ApJS..199...20G/abstract} {199, 20}

\bibitem[\protect\citeauthoryear{{Goldsmith} \& {Langer}}{{Goldsmith} \& {Langer}}{1978}]{goldsmith1978}
{Goldsmith} P.~F.,  {Langer} W.~D.,  1978, \mn@doi [\apj] {10.1086/156206}, \href {https://ui.adsabs.harvard.edu/abs/1978ApJ...222..881G} {222, 881}

\bibitem[\protect\citeauthoryear{{G{\'o}rski}, {Hivon}  \& {et al.,}}{{G{\'o}rski} et~al.}{2005}]{gorski2005}
{G{\'o}rski} K.~M.,  {Hivon} E.,   {et al.,} 2005, \mn@doi [\apj] {10.1086/427976}, \href {https://ui.adsabs.harvard.edu/abs/2005ApJ...622..759G} {622, 759}

\bibitem[\protect\citeauthoryear{{Haardt} \& {Madau}}{{Haardt} \& {Madau}}{2012}]{HaardtMadau2012}
{Haardt} F.,  {Madau} P.,  2012, \mn@doi [\apj] {10.1088/0004-637X/746/2/125}, \href {https://ui.adsabs.harvard.edu/abs/2012ApJ...746..125H} {746, 125}

\bibitem[\protect\citeauthoryear{{Habing}}{{Habing}}{1968}]{habing1968}
{Habing} H.~J.,  1968, \bain, \href {https://ui.adsabs.harvard.edu/abs/1968BAN....19..421H} {19, 421}

\bibitem[\protect\citeauthoryear{{Hanasz}, {Lesch}  \& {et al., }}{{Hanasz} et~al.}{2013}]{Hanasz2013}
{Hanasz} M.,  {Lesch} H.,   {et al., } 2013, \mn@doi [\apjl] {10.1088/2041-8205/777/2/L38}, \href {https://ui.adsabs.harvard.edu/abs/2013ApJ...777L..38H} {777, L38}

\bibitem[\protect\citeauthoryear{{Hanasz}, {Strong}  \& {et al., }}{{Hanasz} et~al.}{2021}]{Hanasz2021}
{Hanasz} M.,  {Strong} A.~W.,   {et al., } 2021, \mn@doi [Living Reviews in Computational Astrophysics] {10.1007/s41115-021-00011-1}, \href {https://ui.adsabs.harvard.edu/abs/2021LRCA....7....2H} {7, 2}

\bibitem[\protect\citeauthoryear{{Heays}, {Bosman}  \& {van Dishoeck}}{{Heays} et~al.}{2017}]{Heays2017}
{Heays} A.~N.,  {Bosman} A.~D.,   {van Dishoeck} E.~F.,  2017, \mn@doi [\aap] {10.1051/0004-6361/201628742}, \href {https://ui.adsabs.harvard.edu/abs/2017A&A...602A.105H} {602, A105}

\bibitem[\protect\citeauthoryear{{Heintz} et~al.,}{{Heintz} et~al.}{2023}]{Heintz2023}
{Heintz} K.~E.,  et~al., 2023, \mn@doi [\apjl] {10.3847/2041-8213/acb2cf}, \href {https://ui.adsabs.harvard.edu/abs/2023ApJ...944L..30H} {944, L30}

\bibitem[\protect\citeauthoryear{{Hillas}}{{Hillas}}{2005}]{Hillas2005}
{Hillas} A.~M.,  2005, \mn@doi [Journal of Physics G Nuclear Physics] {10.1088/0954-3899/31/5/R02}, \href {https://ui.adsabs.harvard.edu/abs/2005JPhG...31R..95H} {31, R95}

\bibitem[\protect\citeauthoryear{{Hollenbach} \& {McKee}}{{Hollenbach} \& {McKee}}{1989}]{Hollenbach1989}
{Hollenbach} D.,  {McKee} C.~F.,  1989, \mn@doi [\apj] {10.1086/167595}, \href {https://ui.adsabs.harvard.edu/abs/1989ApJ...342..306H} {342, 306}

\bibitem[\protect\citeauthoryear{{Hu}, {Naab}  \& {et al., }}{{Hu} et~al.}{2016}]{Hu2016}
{Hu} C.-Y.,  {Naab} T.,   {et al., } 2016, \mn@doi [\mnras] {10.1093/mnras/stw544}, \href {https://ui.adsabs.harvard.edu/abs/2016MNRAS.458.3528H} {458, 3528}

\bibitem[\protect\citeauthoryear{Hunter}{Hunter}{2007}]{Hunter2007}
Hunter J.~D.,  2007, \mn@doi [Computing in Science \& Engineering] {10.1109/MCSE.2007.55}, \href {https://ui.adsabs.harvard.edu/abs/2007CSE.....9...90H/abstract} {9, 90}

\bibitem[\protect\citeauthoryear{{Jaiswal} \& {Omar}}{{Jaiswal} \& {Omar}}{2020}]{Jaiswal2020}
{Jaiswal} S.,  {Omar} A.,  2020, \mn@doi [\mnras] {10.1093/mnras/staa2420}, \href {https://ui.adsabs.harvard.edu/abs/2020MNRAS.498.4745J} {498, 4745}

\bibitem[\protect\citeauthoryear{{Kennicutt}}{{Kennicutt}}{1998}]{Kennicutt1998}
{Kennicutt} Robert~C. J.,  1998, \mn@doi [\apj] {10.1086/305588}, \href {https://ui.adsabs.harvard.edu/abs/1998ApJ...498..541K} {498, 541}

\bibitem[\protect\citeauthoryear{{Kim}, {Kim}, {Gong}  \& {Ostriker}}{{Kim} et~al.}{2023}]{Kim2023}
{Kim} C.-G.,  {Kim} J.-G.,  {Gong} M.,   {Ostriker} E.~C.,  2023, \mn@doi [\apj] {10.3847/1538-4357/acbd3a}, \href {https://ui.adsabs.harvard.edu/abs/2023ApJ...946....3K} {946, 3}

\bibitem[\protect\citeauthoryear{{Kim} et~al.,}{{Kim} et~al.}{2024}]{Kim2024}
{Kim} C.-G.,  et~al., 2024, \mn@doi [arXiv e-prints] {10.48550/arXiv.2405.19227}, \href {https://ui.adsabs.harvard.edu/abs/2024arXiv240519227K} {p. arXiv:2405.19227}

\bibitem[\protect\citeauthoryear{{Klessen}}{{Klessen}}{2000}]{Klessen2000}
{Klessen} R.~S.,  2000, \mn@doi [\apj] {10.1086/308854}, \href {https://ui.adsabs.harvard.edu/abs/2000ApJ...535..869K} {535, 869}

\bibitem[\protect\citeauthoryear{{Krumholz}}{{Krumholz}}{2012}]{Krumholz2012}
{Krumholz} M.~R.,  2012, \mn@doi [\apj] {10.1088/0004-637X/759/1/9}, \href {https://ui.adsabs.harvard.edu/abs/2012ApJ...759....9K} {759, 9}

\bibitem[\protect\citeauthoryear{{Krumholz}, {McKee}  \& {Tumlinson}}{{Krumholz} et~al.}{2009}]{Krumholz2009c}
{Krumholz} M.~R.,  {McKee} C.~F.,   {Tumlinson} J.,  2009, \mn@doi [\apj] {10.1088/0004-637X/699/1/850}, \href {https://ui.adsabs.harvard.edu/abs/2009ApJ...699..850K} {699, 850}

\bibitem[\protect\citeauthoryear{{Krymskii}}{{Krymskii}}{1977}]{Krymskii1977}
{Krymskii} G.~F.,  1977, Akademiia Nauk SSSR Doklady, \href {https://ui.adsabs.harvard.edu/abs/1977DoSSR.234.1306K} {234, 1306}

\bibitem[\protect\citeauthoryear{{Leitherer} et~al.,}{{Leitherer} et~al.}{1999}]{Leitherer1999}
{Leitherer} C.,  et~al., 1999, \mn@doi [\apjs] {10.1086/313233}, \href {https://ui.adsabs.harvard.edu/abs/1999ApJS..123....3L} {123, 3}

\bibitem[\protect\citeauthoryear{{Lelli}, {Verheijen}, {Fraternali}  \& {Sancisi}}{{Lelli} et~al.}{2012}]{Lelli2012}
{Lelli} F.,  {Verheijen} M.,  {Fraternali} F.,   {Sancisi} R.,  2012, \mn@doi [\aap] {10.1051/0004-6361/201117867}, \href {https://ui.adsabs.harvard.edu/abs/2012A&A...537A..72L} {537, A72}

\bibitem[\protect\citeauthoryear{{Markov}, {Carniani}  \& {et al., }}{{Markov} et~al.}{2022}]{Markov2022}
{Markov} V.,  {Carniani} S.,   {et al., } 2022, \mn@doi [\aap] {10.1051/0004-6361/202243336}, \href {https://ui.adsabs.harvard.edu/abs/2022A&A...663A.172M} {663, A172}

\bibitem[\protect\citeauthoryear{{Mather}, {Fixsen}  \& {et al., }}{{Mather} et~al.}{1999}]{Mather1999}
{Mather} J.~C.,  {Fixsen} D.~J.,   {et al., } 1999, \mn@doi [\apj] {10.1086/306805}, \href {https://ui.adsabs.harvard.edu/abs/1999ApJ...512..511M} {512, 511}

\bibitem[\protect\citeauthoryear{{Meyer}}{{Meyer}}{2024}]{Meyer2024}
{Meyer} D.~M.~A.,  2024, \mn@doi [\mnras] {10.1093/mnras/stae870}, \href {https://ui.adsabs.harvard.edu/abs/2024MNRAS.530..539M} {530, 539}

\bibitem[\protect\citeauthoryear{{Nava} \& {Gabici}}{{Nava} \& {Gabici}}{2013}]{Nava2013}
{Nava} L.,  {Gabici} S.,  2013, \mn@doi [\mnras] {10.1093/mnras/sts450}, \href {https://ui.adsabs.harvard.edu/abs/2013MNRAS.429.1643N} {429, 1643}

\bibitem[\protect\citeauthoryear{Nelson \& Langer}{Nelson \& Langer}{1997}]{Nelson_1997}
Nelson R.~P.,  Langer W.~D.,  1997, \mn@doi [ApJ] {10.1086/304167}, \href {https://ui.adsabs.harvard.edu/abs/1997ApJ...482..796N/abstract} {482, 796}

\bibitem[\protect\citeauthoryear{{Neufeld} \& {Wolfire}}{{Neufeld} \& {Wolfire}}{2017}]{Neufeld2017}
{Neufeld} D.~A.,  {Wolfire} M.~G.,  2017, \mn@doi [\apj] {10.3847/1538-4357/aa6d68}, \href {https://ui.adsabs.harvard.edu/abs/2017ApJ...845..163N} {845, 163}

\bibitem[\protect\citeauthoryear{{Omukai}, {Tsuribe}, {Schneider}  \& {Ferrara}}{{Omukai} et~al.}{2005}]{Omukai2005}
{Omukai} K.,  {Tsuribe} T.,  {Schneider} R.,   {Ferrara} A.,  2005, \mn@doi [\apj] {10.1086/429955}, \href {https://ui.adsabs.harvard.edu/abs/2005ApJ...626..627O} {626, 627}

\bibitem[\protect\citeauthoryear{Osterbrock}{Osterbrock}{1988}]{Osterbrock_1988}
Osterbrock D.~E.,  1988, \mn@doi [\pasp] {10.1086/132188}, \href {https://ui.adsabs.harvard.edu/abs/1988PASP..100..412O/abstract} {100, 412}

\bibitem[\protect\citeauthoryear{{Padovani}, {Galli}  \& {Glassgold}}{{Padovani} et~al.}{2009}]{Padovani2009}
{Padovani} M.,  {Galli} D.,   {Glassgold} A.~E.,  2009, \mn@doi [\aap] {10.1051/0004-6361/200911794}, \href {https://ui.adsabs.harvard.edu/abs/2009A&A...501..619P} {501, 619}

\bibitem[\protect\citeauthoryear{{Padovani} et~al.,}{{Padovani} et~al.}{2020}]{PadovaniEtAl2020}
{Padovani} M.,  et~al., 2020, \mn@doi [\ssr] {10.1007/s11214-020-00654-1}, \href {https://ui.adsabs.harvard.edu/abs/2020SSRv..216...29P} {216, 29}

\bibitem[\protect\citeauthoryear{{Padovani} et~al.,}{{Padovani} et~al.}{2022}]{Padovani2022}
{Padovani} M.,  et~al., 2022, \mn@doi [\aap] {10.1051/0004-6361/202142560}, \href {https://ui.adsabs.harvard.edu/abs/2022A&A...658A.189P} {658, A189}

\bibitem[\protect\citeauthoryear{{Palla}, {Salpeter}  \& {Stahler}}{{Palla} et~al.}{1983}]{Palla1983}
{Palla} F.,  {Salpeter} E.~E.,   {Stahler} S.~W.,  1983, \mn@doi [\apj] {10.1086/161231}, \href {https://ui.adsabs.harvard.edu/abs/1983ApJ...271..632P} {271, 632}

\bibitem[\protect\citeauthoryear{Perez \& Granger}{Perez \& Granger}{2007}]{Perez2007}
Perez F.,  Granger B.~E.,  2007, \mn@doi [Computing in Science \& Engineering] {10.1109/MCSE.2007.53}, \href {https://ui.adsabs.harvard.edu/abs/2007CSE.....9c..21P/abstract} {9, 21}

\bibitem[\protect\citeauthoryear{{Peters} et~al.,}{{Peters} et~al.}{2017}]{silcc4}
{Peters} T.,  et~al., 2017, \mn@doi [\mnras] {10.1093/mnras/stw3216}, \href {https://ui.adsabs.harvard.edu/abs/2007CSE.....9c..21P/abstract} {466, 3293}

\bibitem[\protect\citeauthoryear{{Pfrommer}, {Pakmor}  \& {et al., }}{{Pfrommer} et~al.}{2017}]{Pfrommer2017}
{Pfrommer} C.,  {Pakmor} R.,   {et al., } 2017, \mn@doi [\mnras] {10.1093/mnras/stw2941}, \href {https://ui.adsabs.harvard.edu/abs/2017MNRAS.465.4500P} {465, 4500}

\bibitem[\protect\citeauthoryear{{Pineda} et~al.,}{{Pineda} et~al.}{2024}]{Pineda2024}
{Pineda} J.~E.,  et~al., 2024, \mn@doi [arXiv e-prints] {10.48550/arXiv.2402.16202}, \href {https://ui.adsabs.harvard.edu/abs/2024arXiv240216202P} {p. arXiv:2402.16202}

\bibitem[\protect\citeauthoryear{{Rathjen} et~al.,}{{Rathjen} et~al.}{2021}]{silcc6}
{Rathjen} T.-E.,  et~al., 2021, \mn@doi [\mnras] {10.1093/mnras/stab900}, \href {https://ui.adsabs.harvard.edu/abs/2021MNRAS.504.1039R/abstract} {504, 1039}

\bibitem[\protect\citeauthoryear{{Rathjen}, {Naab}  \& {et al., }}{{Rathjen} et~al.}{2023}]{silcc7}
{Rathjen} T.-E.,  {Naab} T.,   {et al., } 2023, \mn@doi [\mnras] {10.1093/mnras/stad1104}, \href {https://ui.adsabs.harvard.edu/abs/2023MNRAS.522.1843R/abstract} {522, 1843}

\bibitem[\protect\citeauthoryear{{Rathjen}, {Walch}  \& {et al.}}{{Rathjen} et~al.}{2024}]{silcc8}
{Rathjen} T.-E.,  {Walch} S.,   {et al.} 2024 (\mn@eprint {arXiv} {2410.00124}), \url {https://arxiv.org/abs/2410.00124}

\bibitem[\protect\citeauthoryear{{R{\'e}my-Ruyer} et~al.,}{{R{\'e}my-Ruyer} et~al.}{2014}]{Remy-Ruyer2014}
{R{\'e}my-Ruyer} A.,  et~al., 2014, \mn@doi [\aap] {10.1051/0004-6361/201322803}, \href {https://ui.adsabs.harvard.edu/abs/2014A&A...563A..31R} {563, A31}

\bibitem[\protect\citeauthoryear{{Ruszkowski} \& {Pfrommer}}{{Ruszkowski} \& {Pfrommer}}{2023}]{Ruszkowski2023}
{Ruszkowski} M.,  {Pfrommer} C.,  2023, \mn@doi [\aapr] {10.1007/s00159-023-00149-2}, \href {https://ui.adsabs.harvard.edu/abs/2023A&ARv..31....4R} {31, 4}

\bibitem[\protect\citeauthoryear{{Sabatini} et~al.,}{{Sabatini} et~al.}{2020}]{Sabatini2020}
{Sabatini} G.,  et~al., 2020, \mn@doi [\aap] {10.1051/0004-6361/202039010}, \href {https://ui.adsabs.harvard.edu/abs/2020A&A...644A..34S} {644, A34}

\bibitem[\protect\citeauthoryear{{Sabatini}, {Bovino}  \& {Redaelli}}{{Sabatini} et~al.}{2023}]{Sabatini2023}
{Sabatini} G.,  {Bovino} S.,   {Redaelli} E.,  2023, \mn@doi [\apjl] {10.3847/2041-8213/acc940}, \href {https://ui.adsabs.harvard.edu/abs/2023ApJ...947L..18S} {947, L18}

\bibitem[\protect\citeauthoryear{Salpeter}{Salpeter}{1955}]{salpeter1955}
Salpeter E.~E.,  1955, \mn@doi [\apj] {10.1086/145971}, \href {https://ui.adsabs.harvard.edu/abs/1955ApJ...121..161S/abstract} {121, 161}

\bibitem[\protect\citeauthoryear{{Sandstrom} et~al.,}{{Sandstrom} et~al.}{2012}]{Sandstrom2012}
{Sandstrom} K.~M.,  et~al., 2012, \mn@doi [\apj] {10.1088/0004-637X/744/1/20}, \href {https://ui.adsabs.harvard.edu/abs/2012ApJ...744...20S} {744, 20}

\bibitem[\protect\citeauthoryear{{Schneider} et~al.,}{{Schneider} et~al.}{2012}]{Schneider2012}
{Schneider} N.,  et~al., 2012, \mn@doi [\aap] {10.1051/0004-6361/201118566}, \href {https://ui.adsabs.harvard.edu/abs/2012A&A...540L..11S} {540, L11}

\bibitem[\protect\citeauthoryear{{Slyz}, {Devriendt}, {Bryan}  \& {Silk}}{{Slyz} et~al.}{2005}]{Slyz2005}
{Slyz} A.~D.,  {Devriendt} J. E.~G.,  {Bryan} G.,   {Silk} J.,  2005, \mn@doi [\mnras] {10.1111/j.1365-2966.2004.08494.x}, \href {https://ui.adsabs.harvard.edu/abs/2005MNRAS.356..737S} {356, 737}

\bibitem[\protect\citeauthoryear{{Socci}, {Sabatini}  \& {et al., }}{{Socci} et~al.}{2024}]{Socci2024}
{Socci} A.,  {Sabatini} G.,   {et al., } 2024, \mn@doi [arXiv e-prints] {10.48550/arXiv.2404.15754}, \href {https://ui.adsabs.harvard.edu/abs/2024arXiv240415754S} {p. arXiv:2404.15754}

\bibitem[\protect\citeauthoryear{{Spitzer} \& {Tomasko}}{{Spitzer} \& {Tomasko}}{1968}]{Spitzer1968}
{Spitzer} Lyman J.,  {Tomasko} M.~G.,  1968, \mn@doi [\apj] {10.1086/149610}, \href {https://ui.adsabs.harvard.edu/abs/1968ApJ...152..971S} {152, 971}

\bibitem[\protect\citeauthoryear{{Stone} \& {et al., }}{{Stone} \& {et al., }}{2019}]{Stone2019}
{Stone} E.~C.,  {et al., } 2019, \mn@doi [Nature Astronomy] {10.1038/s41550-019-0928-3}, \href {https://ui.adsabs.harvard.edu/abs/2019NatAs...3.1013S} {3, 1013}

\bibitem[\protect\citeauthoryear{{Strong}, {Moskalenko}  \& {Ptuskin}}{{Strong} et~al.}{2007}]{Strong2007}
{Strong} A.~W.,  {Moskalenko} I.~V.,   {Ptuskin} V.~S.,  2007, \mn@doi [Annual Review of Nuclear and Particle Science] {10.1146/annurev.nucl.57.090506.123011}, \href {https://ui.adsabs.harvard.edu/abs/2007ARNPS..57..285S} {57, 285}

\bibitem[\protect\citeauthoryear{{Swordy}}{{Swordy}}{2001}]{Swordy2001}
{Swordy} S.~P.,  2001, \mn@doi [\ssr] {10.1023/A:1013828611730}, \href {https://ui.adsabs.harvard.edu/abs/2001SSRv...99...85S} {99, 85}

\bibitem[\protect\citeauthoryear{{Sz{\'e}csi}, {Agrawal}, {W{\"u}nsch}  \& {Langer}}{{Sz{\'e}csi} et~al.}{2022}]{szecsi2020}
{Sz{\'e}csi} D.,  {Agrawal} P.,  {W{\"u}nsch} R.,   {Langer} N.,  2022, \mn@doi [\aap] {10.1051/0004-6361/202141536}, \href {https://ui.adsabs.harvard.edu/abs/2022A%26A...658A.125S/abstract} {658, A125}

\bibitem[\protect\citeauthoryear{{Tielens}}{{Tielens}}{2005}]{Tielens2005}
{Tielens} A.~G.~G.~M.,  2005, {The Physics and Chemistry of the Interstellar Medium}

\bibitem[\protect\citeauthoryear{{Turk}, {Smith}  \& {et al., }}{{Turk} et~al.}{2011}]{Turk2011}
{Turk} M.~J.,  {Smith} B.~D.,   {et al., } 2011, \mn@doi [\apjs] {10.1088/0067-0049/192/1/9}, \href {https://ui.adsabs.harvard.edu/abs/2011ApJS..192....9T} {192, 9}

\bibitem[\protect\citeauthoryear{{Vanzella} et~al.,}{{Vanzella} et~al.}{2023}]{Vanzella2023}
{Vanzella} E.,  et~al., 2023, \mn@doi [\aap] {10.1051/0004-6361/202346981}, \href {https://ui.adsabs.harvard.edu/abs/2023A&A...678A.173V} {678, A173}

\bibitem[\protect\citeauthoryear{Virtanen et~al.,}{Virtanen et~al.}{2020}]{scipy}
Virtanen P.,  et~al., 2020, \mn@doi [Nature Methods] {10.1038/s41592-019-0686-2}, \href {https://rdcu.be/b08Wh} {17, 261}

\bibitem[\protect\citeauthoryear{{Walch} et~al.,}{{Walch} et~al.}{2015}]{silcc1}
{Walch} S.,  et~al., 2015, \mn@doi [\mnras] {10.1093/mnras/stv1975}, \href {https://ui.adsabs.harvard.edu/abs/2015MNRAS.454..238W/abstract} {454, 238}

\bibitem[\protect\citeauthoryear{{Webber}}{{Webber}}{1998}]{Webber1998}
{Webber} W.~R.,  1998, \mn@doi [\apj] {10.1086/306222}, \href {https://ui.adsabs.harvard.edu/abs/1998ApJ...506..329W} {506, 329}

\bibitem[\protect\citeauthoryear{{Wolfire}, {Hollenbach}  \& {et al., }}{{Wolfire} et~al.}{1995}]{Wolfire1995}
{Wolfire} M.~G.,  {Hollenbach} D.,   {et al., } 1995, \mn@doi [\apj] {10.1086/175510}, \href {https://ui.adsabs.harvard.edu/abs/1995ApJ...443..152W/abstract} {443, 152}

\bibitem[\protect\citeauthoryear{{Wolfire}, {McKee}  \& {et al., }}{{Wolfire} et~al.}{2003}]{Wolfire2003}
{Wolfire} M.~G.,  {McKee} C.~F.,   {et al., } 2003, \mn@doi [\apj] {10.1086/368016}, \href {https://ui.adsabs.harvard.edu/abs/2003ApJ...587..278W} {587, 278}

\bibitem[\protect\citeauthoryear{{W{\"u}nsch}, {Walch}, {Dinnbier}  \& {Whitworth}}{{W{\"u}nsch} et~al.}{2018}]{wunsch2018}
{W{\"u}nsch} R.,  {Walch} S.,  {Dinnbier} F.,   {Whitworth} A.,  2018, \mn@doi [\mnras] {10.1093/mnras/sty015}, \href {https://ui.adsabs.harvard.edu/abs/2018MNRAS.475.3393W} {475, 3393}

\bibitem[\protect\citeauthoryear{{W{\"u}nsch}, {Walch}  \& {et al., }}{{W{\"u}nsch} et~al.}{2021}]{wunsch2021}
{W{\"u}nsch} R.,  {Walch} S.,   {et al., } 2021, \mn@doi [\mnras] {10.1093/mnras/stab1482}, \href {https://ui.adsabs.harvard.edu/abs/2021MNRAS.505.3730W/abstract} {505, 3730}

\bibitem[\protect\citeauthoryear{{Zwicky}}{{Zwicky}}{1966}]{Zwicky1966}
{Zwicky} F.,  1966, \mn@doi [\apj] {10.1086/148490}, \href {https://ui.adsabs.harvard.edu/abs/1966ApJ...143..192Z} {143, 192}

\bibitem[\protect\citeauthoryear{van~der Walt, Colbert  \& Varoquaux}{van~der Walt et~al.}{2011}]{vanderWalt2011}
van~der Walt S.,  Colbert S.~C.,   Varoquaux G.,  2011, \mn@doi [Computing in Science \& Engineering] {10.1109/MCSE.2011.37}, \href {https://ui.adsabs.harvard.edu/abs/2011CSE....13b..22V/abstract} {13, 22}

\makeatother
\end{thebibliography}






\bsp	
\label{lastpage}
\end{document}